\documentclass[aps,prx,preprint,bibliography,superscriptaddress]{revtex4-1}
\usepackage[utf8]{inputenc}
\usepackage{amsmath}

\usepackage{physics}
\usepackage{makecell}
\usepackage{mathtools}
\usepackage{amsfonts}
\usepackage{amssymb}
\usepackage{graphicx}
\usepackage{wrapfig}
\usepackage{textcomp}
\usepackage{color}
\usepackage[dvipsnames]{xcolor}
\usepackage{multirow}
\usepackage{comment}
\usepackage{gensymb}
\usepackage{adjustbox}
\usepackage{chemformula}
\usepackage{longtable}
\usepackage{tabularx}

\begin{document}

\title{Non-linear Hall Effects: Mechanisms and Materials}

\author{Arka Bandyopadhyay}
\email{arkabandyopa@iisc.ac.in}
\affiliation{Solid State and Structural Chemistry Unit, Indian Institute of Science, Bangalore 560012, India}
\author{Nesta Benno Joseph}
\affiliation{Solid State and Structural Chemistry Unit, Indian Institute of Science, Bangalore 560012, India}
\author{Awadhesh Narayan}
\email{awadhesh@iisc.ac.in}
\affiliation{Solid State and Structural Chemistry Unit, Indian Institute of Science, Bangalore 560012, India}

\date{\today}

\begin{abstract}
This review presents recent breakthroughs in the realm of nonlinear Hall effects, emphasizing central theoretical foundations and recent experimental progress. We elucidate the quantum origin of the second-order Hall response, focusing on the Berry curvature dipole, which may arise in inversion symmetry broken systems. The theoretical framework also reveals the impact of disorder scattering effects on the nonlinear response. We further discuss the possibility of obtaining nonlinear Hall responses beyond the second order. We examine symmetry-based indicators essential for the manifestation of nonlinear Hall effects in time-reversal symmetric crystals, setting the stage for a detailed exploration of theoretical models and candidate materials predicted to exhibit sizable and tunable Berry curvature dipole. We summarize groundbreaking experimental reports on measuring both intrinsic and extrinsic nonlinear Hall effects across diverse material classes. Finally, we highlight some of the other intriguing nonlinear effects, including nonlinear planar Hall, nonlinear anomalous Hall, and nonlinear spin and valley Hall effects. We conclude with an outlook on pivotal open questions and challenges, marking the trajectory of this rapidly evolving field.
\end{abstract}

\maketitle


\clearpage 

\section{Introduction}

The Hall effect is a well-established phenomenon that occurs when an electric current flows through a conductor placed in a magnetic field~\cite{hall1879new}. 
Discovered by Edwin Hall in 1879, the Hall effect results in the generation of a voltage perpendicular to both the direction of the current and the magnetic field. This transverse voltage, known as the Hall voltage, is proportional to the strength of the magnetic field, the current flowing through the conductor, and a material-specific parameter called the Hall coefficient. 
The Hall effect is widely utilized in various applications, including the measurement of magnetic fields, as well as in the development of Hall effect sensors and devices for detecting current, position, and motion in electronic systems~\cite{ramsden2011hall,chien2013hall}. A generalization in the absence of a magnetic field was also discovered by Hall. 
This is the anomalous Hall effect, where a Hall voltage is generated in a material even in the absence of an external magnetic field~\cite{hall1880rotational}. The anomalous Hall effect typically arises in ferromagnetic materials due to the interaction between the electron spins and the crystal lattice. 
This interaction introduces an additional contribution to the Hall conductivity, resulting in the anomalous Hall voltage. 
The anomalous Hall effect can provide insights into the magnetic properties of materials and is often used as a tool for studying the strength and nature of magnetic interactions in solid-state systems.

The quantum Hall effect is a remarkable manifestation of quantum mechanics in a two-dimensional electron system subjected to low temperatures and a strong magnetic field~\cite{cage2012quantum,huckestein1995scaling,von2005developments}. 
Discovered by Klaus von Klitzing in 1980, nearly a century after Hall's discovery, it revealed that under these conditions, the Hall resistance becomes quantized -- it takes on discrete values that are related to the fundamental constants of nature, such as the elementary charge and Planck's constant. This quantization is incredibly precise and robust, remaining constant even when the sample dimensions or impurities change~\cite{cage2012quantum}. 
The discovery of the quantum Hall effect has had profound implications for the understanding of condensed matter physics, leading to the definition of a new fundamental physical constant and the development of the field of topological insulators and their potential applications in quantum computing~\cite{maciejko2011quantum}. The quantum anomalous Hall effect, on the other hand, involves the generation of a quantized Hall conductance without the need for an external magnetic field~\cite{chang2023colloquium}. 
This effect arises in certain topological or magnetic materials with strong spin-orbit coupling and broken time-reversal symmetry. In a quantum anomalous Hall system, the material becomes an insulator in its bulk but exhibits dissipationless, chiral edge states that contribute to a quantized Hall conductance.

As we can see from the above discussion, traditionally, Hall effect has been tied to the breaking of the time-reversal symmetry, either due to an external magnetic field or the intrinsic magnetism of the material. 
In the past few years, a growing understanding has been developing that a non-linear Hall (NLH) effect may occur in time-reversal symmetric systems~\cite{du2021nonlinear}. The key insight has been that time-reversal symmetry breaking is not essential if the Hall response is considered to non-linear orders and the system has broken inversion symmetry~\cite{sodemann2015quantum}. 
Beginning with pioneering theoretical predictions, there has been a rapidly growing list of materials that have been identified as suitable candidates for exhibiting the NLH effect. Remarkably, a number of recent experiments have discovered such an effect in a diverse variety of materials. Notably, potential applications of the NLH effect have also been recently proposed.
We note that there are two excellent complementary reviews summarizing the early work on the NLH effect~\cite{du2021nonlinear,ortix2021nonlinear}. In this contribution, we review the current state-of-the-art of the field, focusing on the recent progress in mechanisms and materials for the NLH effect. Recently, a new type of linear-response Hall effect has been proposed which is driven by Berry curvature and an inherent electric field in mesoscopic systems with preserved time-reversal symmetry, namely Magnus Hall Effect \cite{papaj2019magnus,sekh2022magnus,mandal2020magnus}. 

In this review, we detail the recent advances in the field of NLH effects. 
We begin with the description of the theoretical formalism of Berry curvature dipole (BCD), having its origins in the Berry curvature. We discuss the second and higher order Hall response, and its connection to nonlinear optical response. This section also discusses the contribution of disorder scattering towards the extrinsic NLH response. 
We next present the symmetry-based indicators required for the emergence of NLH effects under time-reversal symmetric conditions. 
Subsequently, we highlight the different theoretical models and candidate materials predicted to display non-vanishing, tunable BCD. 
The next section discusses the different experimental advances towards the measurement of BCD and NLH response (both intrinsic and extrinsic) in different classes of materials. 
Our final section outlines the most recent developments in the field of non-linear responses beyond NLH, such as the non-linear planar Hall, non-linear anomalous Hall, non-linear spin and valley Hall effects.
We close with an outlook of the important open questions and challenges in this rapidly growing field.

\begin{figure*}[t]
    \centering
    \includegraphics[width=12cm]{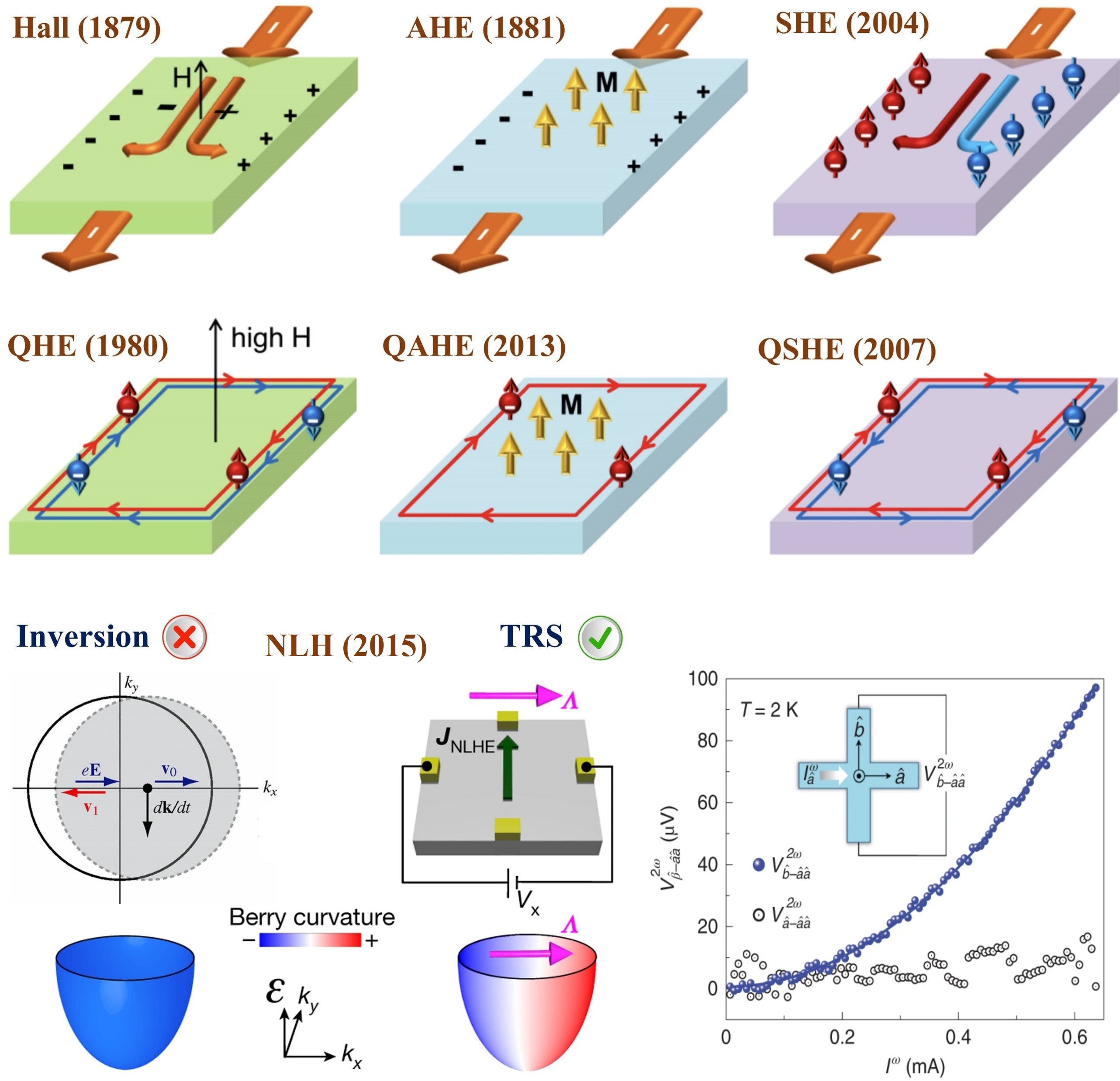}
    \caption{\textbf{Schematic representation of different members of the Hall family.} A comparison of nonlinear Hall (NLH) effect  (bottom panel) with different types of Hall responses -- Hall effect, anomalous Hall effect (AHE), spin Hall effect (SHE), quantum Hall effect (QHE), quantum anomalous Hall effect (QAHE) and quantum spin Hall effect (QSHE). The corresponding year of discovery is indicated in parentheses. Here, $H$, $M$, and $\Lambda$ represent magnetic field strength, magnetization, and Berry curvature dipole magnitude, respectively. Under time-reversal symmetry, a nonuniform distribution of Berry curvature can be achieved by external perturbations, such as an electric field, that leads to the non-vanishing nonlinear Hall response in inversion symmetry broken systems. Adapted from Refs.~\cite{ma2019observation,moore2010confinement,chang2016quantum,kumar2021room}.}
    \label{fig:firstfigure}
\end{figure*}

\section{Theoretical background}

The foundational work by Sodemann and Fu~\cite{sodemann2015quantum}, leveraging insights derived from a seminal study by Moore and Orenstein~\cite{moore2010confinement}, has led to the formulation of the NLH effect. The NLH effect stands as the latest member of the Hall family, and its distinct origin sets it apart from other members, as illustrated in Fig.~\ref{fig:firstfigure}. In particular, a connection between NLH responses and a newly introduced geometric quantity, BCD, has been established, as we will discuss. Recently, Matsyshyn and Sodemann~\cite{matsyshyn2019nonlinear} have introduced an alternative elegant perspective on BCD that occurs in inversion symmetry broken metals. In particular, BCD has been projected as a naturally occurring non-Newtonian Drude weight that invariably acts as an order parameter of metals without inversion symmetry. This is exciting as the search for a measurable order parameter for the inversion symmetry broken metals was a long-existing research question. In this section, we will discuss the quantum origin of the BCD in metals, followed by the more usual semiclassical treatment. The two complementary views, give direct physical insights into these phenomena. 

For this purpose, let us first write down the position operator $(\hat{r}_{\alpha \beta}$) in the Bloch basis~\cite{book}, as follows

\begin{equation}
    \hat{\vec{r}}_{\alpha \beta} = \delta_{\alpha \beta} \: i \:\hat{\partial}_{\vec{k}} + \hat{A}_{\alpha \beta} (\vec{k}).
    \label{eq:equation}
\end{equation}

Here the non-Abelian Berry connection, $\hat{A}_{\alpha \beta} (\vec{k})$, can be expressed in terms of periodic Bloch functions as $\hat{A}_{\alpha \beta} (\vec{k}) = i \: \langle \alpha k | \partial_k | \beta k \rangle$, where \{$\alpha$, $\beta$\} are band indices and $\vec{k}$ represents the crystal momentum. Using the position operator mentioned above, it is straightforward to derive the Hamiltonian under a time-varying electric field, $\vec{\mathcal{E}}(t)$, that is otherwise spatially uniform

\begin{equation}
    \hat{\mathcal{H}} = \hat{\mathcal{H}}_0 + q \: \hat{\vec{r}} \cdot \vec{\mathcal{E}}(t).
    \label{eq:hamiltonian}
\end{equation}

The above description is given in the so-called ``length gauge", where the interaction term is defined by $\hat{\vec{r}} \cdot \vec{\mathcal{E}} (t)$. On the other hand, the ``velocity gauge" is underpinned by the conventional minimal substitution scheme, $\hat{\mathcal{H}}(\vec{k}) \rightarrow  \hat{\mathcal{H}}\left(\vec{k} + e \vec{A}(t)\right)$. The vector potential like term, $\vec{A}(t)$, is related to the electric field by the relation $\vec{\mathcal{E}} = - \partial_{t} \vec{A}(t)$ and contributes to the interaction term as $-i \Vec{A}(t) \cdot \vec{\nabla}$. These two gauges have different advantages and drawbacks depending on the problem. The length gauge is generally used to obtain analytical solutions, sometimes in semiclassical approximations for various optical and transport properties, including topological ones. On the other hand, the velocity gauge is better suited for numerical methods, including the tight-binding approach~\cite{konig2019gyrotropic,isobe2020high,du2019disorder,xiao2019theory}. A comparison of these two widely used gauges is given in Table~\ref{tab:gauge}. Once the Berry connection is calculated, the Hamiltonian given in Eq.~\ref{eq:hamiltonian} can be projected to a single band. We note that the commutators of the position operator projected on a given band along different directions can be written as

\begin{equation}
    \left[\hat{r}_{\alpha}^{\mu} , \hat{r}_{\alpha}^{\nu} \right] = i ( \partial_{k^{\mu}} \hat{A}_{\alpha}^{\nu} - \partial_{k^{\nu}} \hat{A}_{\alpha}^{\mu}) = i \: \Omega_{\alpha}^{\mu \nu}.
    \label{eq:curl}
\end{equation}

The above equation reveals the non-commuting nature of different components of the projected position operator, which essentially manifests the Berry curvature of the system. In other words, the commutator is equivalent to executing a derivative operation ($\vec{\nabla}$) on the band diagonal elements of the Berry connection matrix. As a consequence, we have obtained the Berry curvature -- an anti-symmetric second-rank tensor -- which appears to be the topological property of a single band. However, the concept of Berry curvature strongly depends on the virtual transitions between different bands when the velocity operator acts as an inter-band coupling parameter~\cite{hasan2010colloquium,cayssol2021topological}. We note that the tensor and pseudovector form of the Berry curvature is related by the expression  $\Omega _{{\alpha,\mu \nu }}=\epsilon _{{\mu \nu \lambda }}\,{\Omega }_{{\alpha,\lambda }}$. Here $\epsilon _{{\mu \nu \lambda }}$ is the Levi-Civita tensor. Moreover, the velocity operator can be obtained by calculating the time rate of change of the position operator as

\begin{equation}
    \hat{v}^{\mu}_{\alpha}  = \frac{i}{\hbar} \left[ \hat{\mathcal{H}}(k) ,  \hat{r}^{\mu}  \right] = \frac{1}{\hbar}\: \partial_{k^{\mu}} \epsilon_\alpha - \frac{e \: \mathcal{E}^{\nu}(t)}{\hbar} \Omega_{\alpha}^{\mu \nu}.
    \label{eq:vel}
\end{equation}

Similarly, we can evaluate the acceleration operator from the rate of change of above velocity operator ($\mu, \nu, \lambda \in \{x,y,z\}$), as given below

\begin{equation}
       \hat{a}^{\lambda}_{\alpha} =  - \frac{e \: \mathcal{E}^{\mu}(t)}{\hbar^2} \: \partial_{k^{\mu}}\partial_{k^{\lambda}} \epsilon_{\alpha} - \frac{e}{\hbar} \: \partial_{t} \mathcal{E}^{\mu} \: \Omega_{\alpha}^{\mu \lambda} + \frac{e^2}{\hbar^2} \: \mathcal{E}^{\mu} \mathcal{E}^{\nu} \: \partial_{k^{\mu}} \Omega_{\alpha}^{\nu \lambda}.  
       \label{eq:acc}
\end{equation}

The acceleration, in the above Eq.~\ref{eq:acc}, has three distinct terms: (i) the first term is proportional to the force and thus provides the linear Drude weight or inverse inertia using Newtonian mechanics, (ii) the second term depends on the time variation of the electric field and is not significant unless the electric field changes rapidly with time, and (iii) the third term is fascinating as it is not only orthogonal to the external electric field but also proportional to the second order of the electric field in a non-Newtonian fashion. This term is essentially a non-Newtonian acceleration, and the corresponding nonlinear Drude weight is a tensor quantity that depends on the gradient of the Berry curvature in the reciprocal space, also known as the BCD. The above perspective on BCD efficiently helps to differentiate the intrinsic contribution of the NLH effect from all the other extrinsic impurity-induced effects~\cite{konig2019gyrotropic,isobe2020high,du2019disorder,xiao2019theory}. 

\begin{table}[t]
    \centering
     \caption{A comparison between the advantages and disadvantages of two often-used gauges, i.e., length and velocity gauges. The table is adapted from Refs.~\cite{ma2021topology,matsyshyn2019nonlinear,ventura2017gauge}.}
\renewcommand{\arraystretch}{1.6} 
\setlength{\arrayrulewidth}{0.1mm} 
     \begin{tabular}{c|c | c}
     \hline
     \hline
     Gauge & Advantages & Disadvantages \\
\hline
\hline
\multirow{2}{4em}{Length}  & \shortstack{Easy access to \\  the semiclassical approaches} & \shortstack{Numerical differentiation of position \\ operator with momentum is not convenient} \\ 
& \shortstack{Definition of some topological \\ properties become straightforward} & \shortstack{Frequency poles of higher order \\ should be taken into account} \\
\hline
\multirow{2}{4em}{Velocity} & \shortstack{Feynman diagram method \\ becomes convenient} & \shortstack{Unreliable results because of \\ truncated summation of bands} \\
& \shortstack{Absence of momentum differentiation \\ makes it numerically convenient } & \shortstack{Low-energy divergences sometimes \\ need to be dealt with} \\
\hline
\hline
    \end{tabular}
      \label{tab:gauge}
\end{table}

 \begin{figure*}[t]
    \centering
    \includegraphics[width=12 cm]{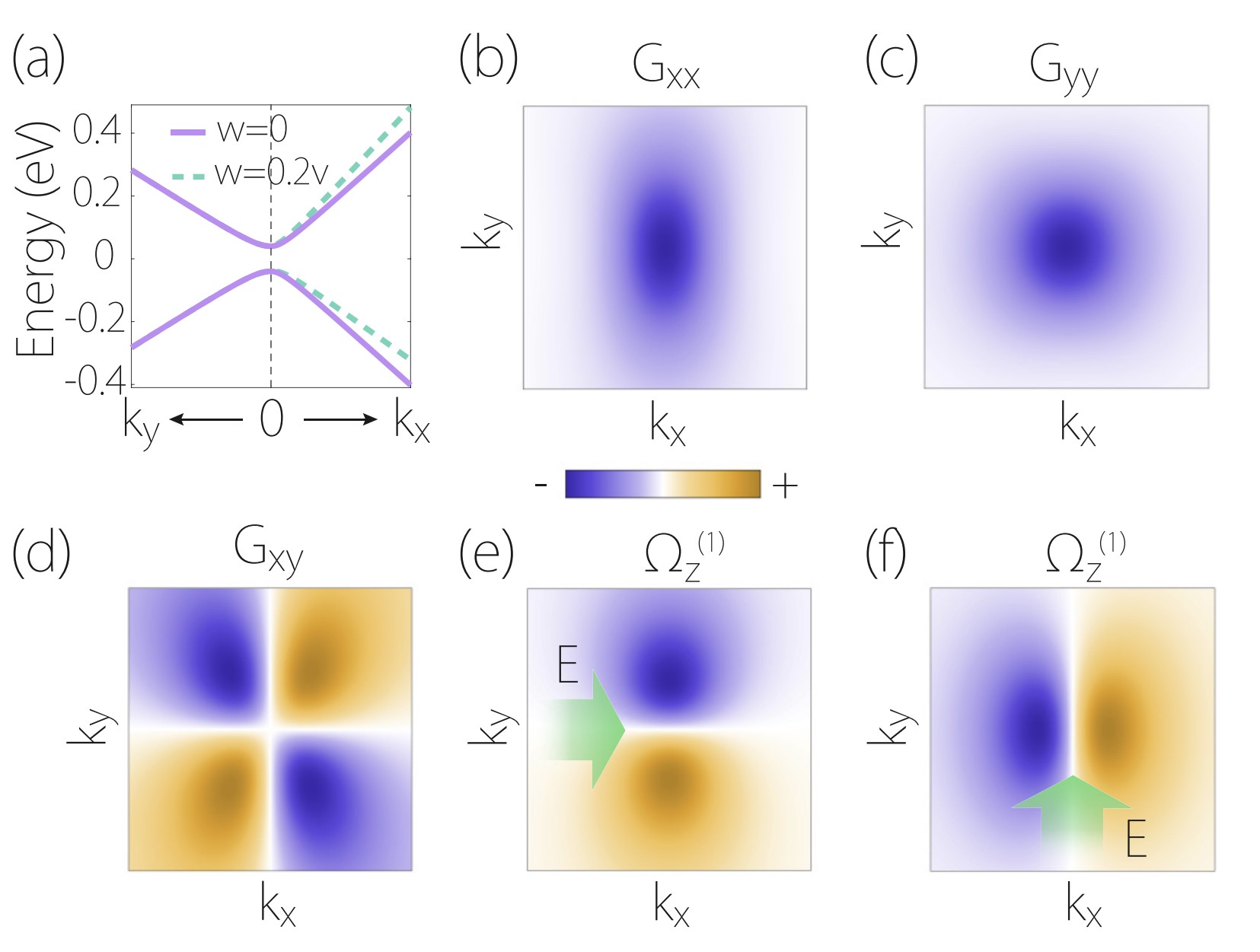}
    \caption{\textbf{Band structure and Berry connection polarizability tensor components.} (a) The electronic band structure of the 2D Dirac model is shown along the $k_x$ and $k_y$ directions. The solid lines indicate the scenario without the tilt term, whereas the dashed lines represent the case with the tilt. (b)–(d) Illustrations of the distribution of Berry curvature polarizability tensor components for the valence band. (e)-(f) The field-induced Berry curvature when an electric field, marked by the green arrow, is applied along the (e) $x$-direction and (f) $y$-direction. In the calculation, the parameters are set as follows: $v_x = 1 \times 10^6$ m/s, $v_y = 0.7 v_x$, $t = 0.2 v_x$, and $\Delta = 40$ meV. Reproduced with permission from~\cite{liu2022berry}.}
    \label{fig:higherorder}
    \end{figure*}

\begin{figure*}[t]
    \centering
    \includegraphics[width=12cm]{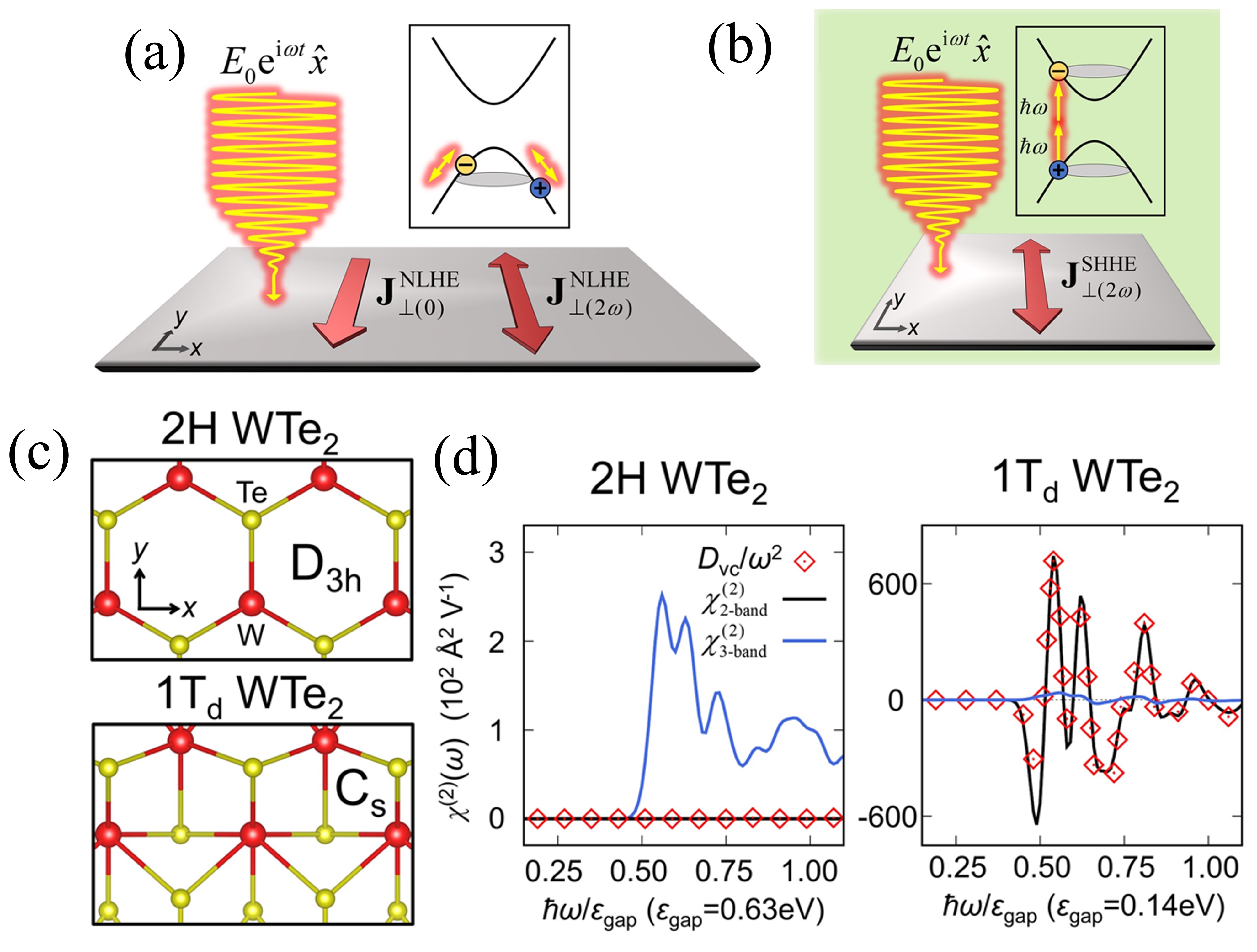}
    \caption{\textbf{The interband BCD and second harmonic generation in insulators.} Schematic representation of the (a) NLH response and (b) second harmonic Hall effect observed in the metallic and insulating systems, respectively. The systems are exposed to electromagnetic waves with frequency $\omega$, which is linearly polarized along the $x$-direction. In the case of the NLH effect, an alternating current with frequency $2 \omega$ is obtained in addition to a rectified direct current. However, in the insulating systems, the incident wave triggers electron-hole generation, resulting in an alternating current with frequency $2 \omega$ in the orthogonal direction to the polarization. (c) The structure of the \ch{WTe2} system in the 2H phase (top panel) and 1T$_d$ phase (bottom panel). The 2H and 1T$_d$ phases exhibit $D_{3h}$ and $C_s$ point groups, respectively. (d) The corresponding second harmonic susceptibility for 2H (left panel) and 1T$_d$ (right panel). The interband BCD ($D_{vc}/\omega^2$) is in excellent agreement with the two-band transition term in both cases and has an appreciable contribution when the crystal symmetry is reasonably low. Reproduced with permission from Ref.~\cite{okyay2022second}.}
    \label{fig:okyay2022second}
\end{figure*}

In the above discussion, we have obtained different linear and nonlinear contributions to the Hall acceleration using a quantum analytical formalism. Another straightforward way to understand these distinct sources of the current density is to utilize a semiclassical approach as introduced by Sodemann and Fu~\cite{sodemann2015quantum}. The electric current density ($j_{\alpha}$) can be determined by integrating the physical velocity ($v_{\alpha}$) of electrons, taking into account their corresponding occupation probability, as given by,

\begin{equation}
    j_{\alpha} = -e \int_{k} f(\epsilon) v_{\alpha}. 
    \label{eq:currentdensity}
\end{equation}

In the above Eq.~\ref{eq:currentdensity}, $\int_{k} = \int d^{d} k / (2 \pi)^d$ and the velocity $v_\alpha$ has two distinct contributions: (i) group velocity of the electrons with dispersion $\epsilon(k)$ and (ii) the anomalous velocity caused by Berry curvature ($\Omega$), written as,

\begin{equation}
    v_{\alpha} = \partial_{\alpha} \epsilon(k) + \epsilon_{\alpha \beta \gamma} \Omega_{\beta} \dot{k}_{\gamma}.
\end{equation}

The velocity operator mentioned above is same as Eq.~\ref{eq:vel}, when $\dot{k}_{\gamma}$ is expressed as $-e \mathcal{E}_\gamma (t)$ ($\hbar = 1$). On the other hand, the distribution function, $f(\epsilon)$, can be obtained by solving the Boltzmann transport equations in the presence of an external electric field,

\begin{equation}
    f(\epsilon) = f_{0} (\epsilon) + e \tau \: \vec{v}_{0} \cdot \vec{\mathcal{E}} \:\partial_{\epsilon}f_{0}(\epsilon) + ...,
\end{equation}

where $\vec{v}_{0}$ and $\tau$ represent the band velocity and the relaxation time, respectively. Substituting the expressions of the velocity and distribution function in the current density expression, we obtain 

\begin{equation}
    j_{\mu} = \sigma_{\mu \nu} \mathcal{E}_{\nu} + \xi_{\mu \nu \lambda} \mathcal{E}_{\nu} \mathcal{E}_{\lambda}.
    \label{eq:conductivity}
\end{equation}

The first term $\sigma_{\mu \nu}$ is the linear conductivity term, which contains one symmetric part (Ohmic type) that depends on the band velocity and another antisymmetryic part (Hall type) that depends on the Berry curvature. This reads

\begin{equation}
    \sigma_{\mu \nu} = - e^2 \tau \int_{k} v_{\mu} v_{\nu} \partial_{\epsilon}f_{0}(\epsilon) - \frac{e^2}{h} \int_{k} \Omega_{\mu \nu} f_{0}(\epsilon).
\end{equation}

On the other hand, the second term, $\xi_{\mu \nu \lambda}$, represents a quadratic conductivity tensor, which depends on the first-order moment of the Berry curvature -- responsible for the NLH effect. It can be written as

\begin{equation}
    \xi_{\mu \nu \lambda} = - \frac{e^3 \tau}{\hbar} \int_{k} \Omega_{\mu \nu} v_{\lambda} \partial_{\epsilon}f_{0}(\epsilon) = \frac{e^3 \tau}{\hbar^2} \int_{k} \partial_{\lambda} \Omega_{\nu \lambda} f_{0}(\epsilon) = - \xi_{\nu \mu \lambda}.
    \label{eq:nheexpress}
\end{equation}

In the above Eq.~\ref{eq:nheexpress}, the presence of partial differentiation of the distribution function, $\partial_{\epsilon}f_{0}(\epsilon)$, reveals that the states near the Fermi surface of the system will have a predominant contribution to the nonlinear conductivity at low temperature~\cite{haldane2004berry}. 

We can understand the above formalism also from the Boltzmann equation framework governing the electron distribution under the assumption of constant relaxation time ($\tau$) as below~\cite{mahan2000many},

\begin{equation}
    e \tau \mathcal{E}_{\alpha} \partial_{\alpha} f - \tau \partial_{t} f = f - f_{0}.
\end{equation}

Here the unperturbed distribution function is represented by $f_0$. Now, as our goal is to compute the response to the second order in the electric field, we undertake the expansion of the distribution up to the second order, represented by $f=$Re\{$f_0$ + $f_1$ + $f_2$\}, where

\begin{equation}
    f_1 = \left[\frac{e \tau \mathcal{E}_{\alpha} \partial_{\alpha} f_{0}}{1 + i \omega \tau} \right] e^{i \omega t} \; \; \; \; \textup{and} \; \; \; \;
    f_2 = \frac{e^2 \tau^{2} \mathcal{E}^{*}_{\alpha} \mathcal{E}_{\beta} \partial_{\alpha \beta} f_0}{2 (1+ i \omega \tau)} + \left[\frac{e^2 \tau^{2} \mathcal{E}_{\alpha}\mathcal{E}_{\beta} \partial_{\alpha \beta} f_0}{2(1+i \omega \tau)(1+2i \omega \tau)}\right] e^{2 i \omega t }.  
\end{equation}

Using the above expressions, we can obtain the current density, $j_\alpha = Re\{ j_{\alpha}^{0} + j_{\alpha}^{2\omega} e^{2 i \omega t}\}$, where $j_{\alpha}^{0}$ and $j_{\alpha}^{2\omega}$, have the following form

\begin{align}
    j_{\alpha}^{0} & = \frac{e^2}{2} \int_{k} \varepsilon_{\alpha \beta \gamma} \Omega_{\beta} \mathcal{E}_{\gamma}^{*} \left[ \frac{e \tau \mathcal{E}_{\alpha} \partial_{\alpha} f_{0}}{1 + i \omega \tau}\right] -e \int_{k} \left[ \frac{e^2 \tau^{2} \mathcal{E}^{*}_{\alpha} \mathcal{E}_{\beta} \partial_{\alpha \beta} f_0}{2 (1+ i \omega \tau)} \right] \partial_{\alpha} \epsilon(k), \nonumber \\
     j_{\alpha}^{2\omega} & =  \frac{e^2}{2} \int_{k} \varepsilon_{\alpha \beta \gamma} \Omega_{\beta} \mathcal{E}_{\gamma} \left[ \frac{e \tau \mathcal{E}_{\alpha} \partial_{\alpha} f_{0}}{1 + i \omega \tau}\right]
    - e \int_{k} \left[\frac{e^2 \tau^{2} \mathcal{E}_{\alpha}\mathcal{E}_{\beta} \partial_{\alpha \beta} f_0}{2(1+i \omega \tau)(1+2i \omega \tau)}\right] \partial_{\alpha} \epsilon(k).
    \label{eq:eqns}
\end{align}

We note that the current component $j_{\alpha}^{0}$ signifies the rectified current, while $j_{\alpha}^{2\omega}$ represents the second harmonic. Among different terms in Eq.~\ref{eq:eqns}, only the following terms are nonzero under time-reversal symmetry~\cite{sodemann2015quantum,deyo2009semiclassical} -- $j_{\alpha}^{0} = \chi_{\alpha \beta \gamma} \mathcal{E}_{\beta} \mathcal{E}^{*}_{\gamma}$ and $j_{\alpha}^{2 \omega} = \xi_{\alpha \beta \gamma} \mathcal{E}_{\beta} \mathcal{E}_{\gamma}$, where $\xi_{\alpha \beta \gamma}$ reads as

\begin{equation}
    \xi_{\alpha \beta \gamma} = \varepsilon_{\alpha \delta \gamma} \frac{e^3 \tau}{2(1+i \omega \tau)} \int_{k} \Omega_{\delta} \partial_{\beta} f_{0}.
\end{equation}

The above equation is identical to the one given in Eq.~\ref{eq:nheexpress}. Here the integration $\int_{k} \Omega_{\delta} \partial_{\beta} f_{0}$ gives rise to the second-order current response, which can also be written as 

\begin{equation}
    D_{\alpha \beta} = \int_{k} f_{0} (\partial_{\alpha} \Omega_{\beta})
\label{eq:bcdexp1}.
\end{equation}

The tensor quantity $D_{\alpha \beta}$ as given in Eq.~\ref{eq:bcdexp1} is the "dipole of the Berry curvature" in momentum space, known as the BCD. Now, similar to nonlinear optics~\cite{bloembergen1982nonlinear,autere2018nonlinear}, the emergence of the second-order response in electronic conductivity strictly depends on the symmetry-based indicators. However, unlike nonlinear optical responses (frequency $\sim 10^{14}$ Hz), the frequency regime of the NLH response is significantly lower. In standard practice, these nonlinear responses are estimated using the semiclassical Boltzmann transport equations~\cite{sodemann2015quantum,du2019disorder} or the reduced density matrix formalism~\cite{sipe1993nonlinear,ventura2017gauge}. Both of these approaches invariably adopt the relaxation time approximation to deal with the dissipation processes. However, the relaxation time approximation does not respect the length and velocity gauge invariance~\cite{ventura2017gauge}, and is not well-suited for the generalized forces with non-vanishing frequency~\cite{michishita2021effects}. To overcome the issues mentioned above, Michishita \textit{et al.} have proposed an alternative pathway using Green's function formalism to study the role of dissipation on linear conductivity and beyond~\cite{michishita2022dissipation}. In particular, it has been found that the poles of the Green’s function primarily give rise to the higher-order response in presence of dissipation. Here we discussed the response upto second order. However, recently, higher order Hall response has also been investigated -- we address this next.

\subsection{Higher order Hall response}

In the preceding section, we have discussed that an anomalous Hall effect in magnetic systems probes the notion of Berry curvature~\cite{nagaosa2010anomalous,xiao2010berry}. In a similar vein, detection of the nonlinear or second-order Hall response in time-reversal symmetric systems leads to the concepts of BCD -- another band geometric property. However, it naturally prompts us to inquire -- is it possible to extend our analysis beyond the second order? Theoretically, a connection between the third-order Hall effect in non-magnetic materials and the Berry-connection polarizability tensor has been established~\cite{gao2014field, pal2024polarization}, and very recently experimentally verified in the T$_{d}$ phase of MoTe$_2$ systems~\cite{lai2021third}. Through the single-band Boltzmann transport equation, it has been shown that the third-order nonlinear Hall response persists in systems lacking time-reversal symmetry. This response is directly proportional to the second derivative of the Berry curvature concerning momentum~\cite{parker2019diagrammatic,zhang2023higher}. Gao \textit{et al.} have delved into the theoretical examination of third-harmonic generation within this third-order NLH effect in the presence of an external electromagnetic field by employing quantum kinetic equations~\cite{gao2023third}. In this context, we will discuss the significance of the Berry connection polarizability tensor -- another intrinsic measure of band geometry -- in generating the third-order Hall effect~\cite{liu2022berry}. Let us examine the role of polarizability tensor by employing a prototype two-dimensional Dirac model as given below

\begin{equation}
    H(k) = t k_x \mathbb{I} + v_x k_x \sigma_x + v_y k_y \sigma_y + \Delta \sigma_z.
\label{eq:tightbindingthirdorder}
\end{equation}

Here $\sigma$, the set of Pauli matrices, serve as the basis of the Hamiltonian, and $\mathbb{I}$ is the identity matrix. The other parameters, $t$, $v_{x/y}$, $k_{x/y}$, and $\Delta$ are the tilt parameter, $x/y$ component of the velocity, $x/y$ component of the momentum, and the mass term, respectively. The energy dispersion relation can be obtained as

\begin{equation}
    \epsilon_{\pm} = t k_x \pm \left[ v_{x}^{2}k_{x}^{2} + v_{y}^{2}k_{y}^{2} + \Delta^2 \right]^{1/2}.
\end{equation}

The dispersion relation is plotted in presence and absence of the tilt parameter in Fig.~\ref{fig:higherorder}(a). Using the above, the Berry connection polarizability tensor can be evaluated using

\begin{equation}
    G_{\alpha \beta} = 2 \mathrm{Re} \sum_{m \neq n } \frac{(A_{\alpha})_{nm} (A_{\beta})_{mn}}{\epsilon_{n} - \epsilon_{m}}.
\end{equation}

Here $A$ represents the interband ($n$ and $m$) Berry connection. In case of the Dirac model the Berry connection polarizability tensor has the form

\begin{gather}
 [G]
 =
 - \frac{v_{X}^{2} k_{x}^{2}}{3 \left[ v_{x}^{2}k_{x}^{2} + v_{y}^{2}k_{y}^{2} + \Delta^2 \right]^{5/2}}
  \begin{bmatrix}
  k_{y}^{2} + \Delta^2/v_{y}^{2} &
   - k_{x} k_{y} \\
  - k_{x} k_{y} &
    k_{x}^{2} + \Delta^2/v_{x}^{2} 
   \end{bmatrix}.
   \label{eq:highereq}
\end{gather}

Fig.~\ref{fig:higherorder}(b)-(d) portray the components of the Berry connection polarizability tensor given in Eq.~\ref{eq:highereq}. We note that the components of the tensor remain unaffected by the tilt term. Notably, the diagonal elements of the tensor exhibit a monopole-like structure, reaching a peak at the center where the gap is minimal. On the contrary, the off-diagonal component, $G_{xy}$, displays a quadrupole-like structure. In the presence of an in-plane electric field $\vec{E}$, the Berry curvature can be expressed as

\begin{equation}
    \Omega(\vec{k}) = \frac{v_{x}^{2} v_{y}^{2}}{2 \left[ v_{x}^{2}k_{x}^{2} + v_{y}^{2}k_{y}^{2} + \Delta^2 \right]^{5/2}} \: (\vec{k} \times \vec{E}). 
\end{equation}

Fig.~\ref{fig:higherorder}(e)-(f) reveal a dipole-like structure in the Berry curvature, in contrast to the original monopole structure. Notably, the orientation of the dipole is contingent upon the applied electric field and can be manipulated by rotating the electric field, as depicted in Fig.~\ref{fig:higherorder}(e)-(f). 

The Berry connection polarizability tensor gives rise to the third-order nonlinear Hall effect, which is suppressed for the two-dimensional point groups $C_{3v}$, $C_{6v}$, $D_3$, $D_{3h}$, $D_{3d}$, $D_{6}$, and $D_{6h}$. In this regard, Wei \textit{et al.} have formulated a quantum nonlinear transport theory to explore the third-order Hall response within a four-terminal setup characterized by time-reversal symmetry in the quantum regime~\cite{wei2022quantum}. The emergence of a third-order Hall effect in the surface states of hexagonal topological insulators (e.g., Bi$_2$Te$_3$) has been examined solely under the influence of an electric field~\cite{nag2023third}. It has been found that the tilt and hexagonal warping on the Berry connection polarizability tensor and, consequently, the topological orbital Hall effect triggers a gap closing in Dirac cones. The third-order response is significantly enhanced with increasing tilt strength and warping values. Similarly, a hexagonal warping effect has been incorporated to study the emergence of BCD-induced second-order Hall effect and Berry connection polarizability tensor induced third-order Hall response in Rashba systems~\cite{saha2023nonlinear}. Multi-Weyl semimetals~\cite{huang2016new} are another class of materials to obtain topological charge dependent quantum geometric properties, such as BCD and Berry connection polarizability tensor. Utilizing low-energy effective models, Roy \textit{et al.} have derived comprehensive analytical expressions, unveiling the significance of BCD and Berry connection polarizability tensor in second and third order effects~\cite{roy2022non}. It has further been found that Berry curvature multipoles, representing higher moments of Berry curvatures at the Fermi energy, can give rise to higher-order NLH effects. In particular, an alternating current Hall voltage perpendicular to the current direction emerges, with the frequency being an integer multiple of the applied current frequency. Notably, the symmetry analysis across various three- and two-dimensional magnetic point groups reveals that quadrupole, hexapole, and even higher Berry curvature moments can lead to a prominent frequency multiplication in specific materials~\cite{zhang2023higher}. Very recently, the third-order nonlinear response in two-dimensional altermagnets protected by $C_n T$ symmetry has been discovered~\cite{fang2023quantum}, revealing a dependence on quantum geometry components -— quantum metric and Berry curvature. Longitudinal responses for all planar altermagnets arise from the quantum metric quadrupole, while transverse responses involve both quantum metric quadrupole and Berry curvature quadrupole. Furthermore, the responses are highly sensitive to crystalline anisotropy and weak spin-orbit coupling.

\begin{figure*}[t]
    \centering
    \includegraphics[width=14 cm]{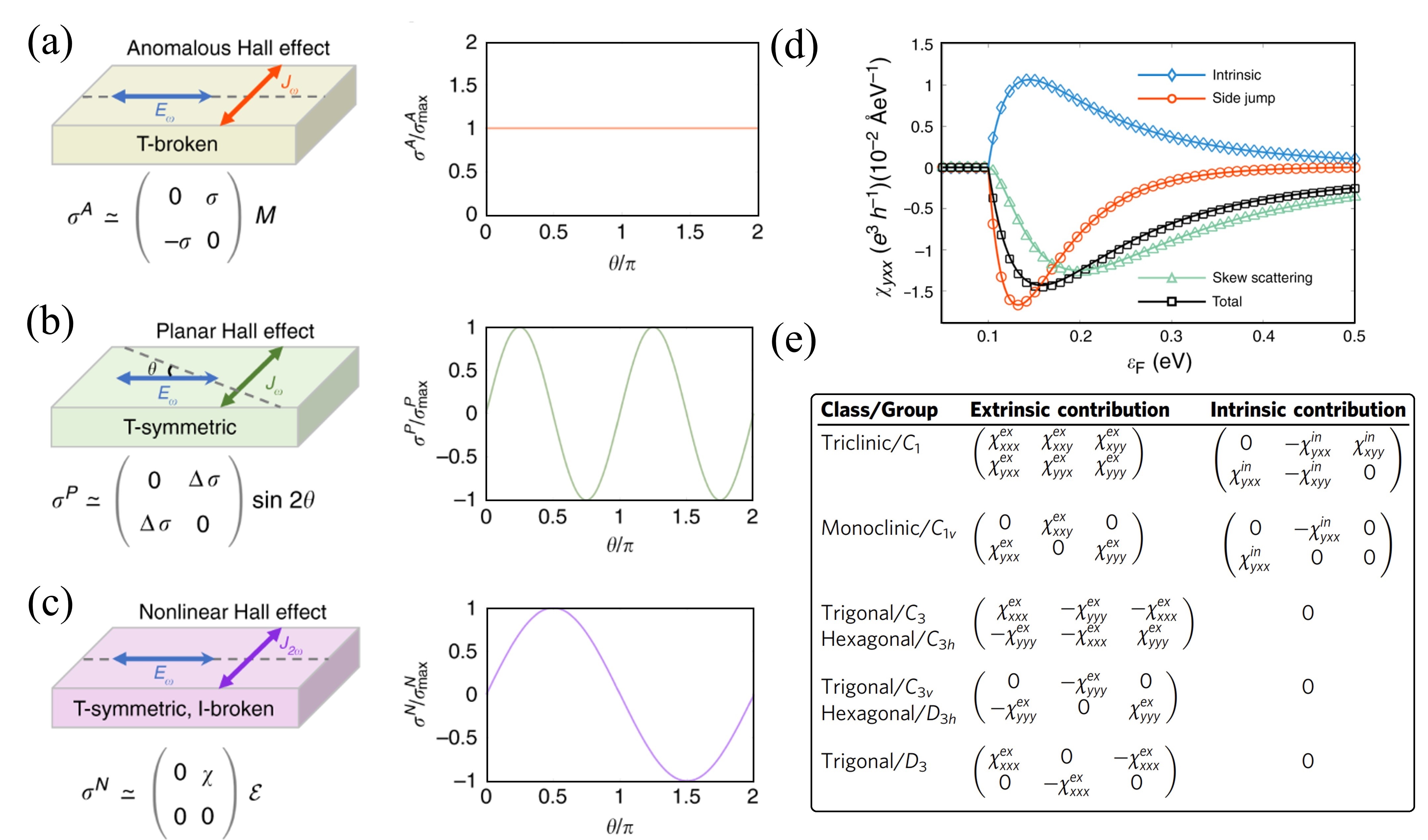}
    \caption{\textbf{Disorder-generated NLH effect.} A comparison between the linear and NLH effects in the systems without external magnetic field. Schematic of the experimental setups for the (a) anomalous, (b) planar, and (c) NLH effects at different time reversal and inversion symmetric conditions. The anomalous Hall conductivity, denoted as $\sigma^A$, is an anti-symmetric tensor. The symbol $M$ denotes the magnetization, while $\sigma^{P}$ stands for planar Hall conductivity. The difference between longitudinal conductivity along the two principal axes is given by $\Delta \sigma \equiv \sigma_{\parallel} - \sigma_{\perp}$. The angle between the electric field ($\mathcal{E}$) and one of the principal axis ($\parallel$) is represented by $\theta$. In the case of the NLH effect, $\sigma^N$ stands for the conductivity with NLH response tensor, $\chi$. The right panels show that the angular dependence efficiently differentiates different types of Hall response. (d) The individual contributions of intrinsic, side-jump, and skew-scattering to the total NLH conductivity, $\xi_{yxx}$, of a generic two-dimensional tilted massive Dirac model. Parameters used are $t$ = 0.1 eV\AA{}, $v$ = 1 eV\AA{}, $m$=0.1 eV, $n_i V_{0}^{2} = 10^2$ eV$^2$\AA{}$^2$, and $n_i V_{1}^{3} = 10^4$ eV$^3$\AA{}$^4$. (e) The non-zero elements of the NLH conductivity tensor for 32 distinct point groups in a two-dimensional systems. The superscript `in' and `ex' are the intrinsic and extrinsic contributions. Reproduced with permission from~\cite{du2019disorder,du2021quantum}.}
    \label{fig:intrinsicextrinsic}
\end{figure*}

\subsection{Analogy with non-linear optical response}

In the above discussion, we have established BCD as a Fermi surface property that essentially behaves like a non-Newtonian Drude weight. Therefore, metallic systems are the usual platforms to realize the BCD-induced effects, such as NLH responses. However, the interband BCD~\cite{okyay2022second}, an analog of the BCD derived for metals (Fig.~\ref{fig:okyay2022second}(a)), can be defined for systems with a finite band gap, as follows    

\begin{equation}
    D_{vc} = \int_{k} \partial_{k} \Omega_{vc} \: \Theta(2 \omega - \omega_{cv}).
    \label{eq:intraBCD}
\end{equation}

In the above Eq.~\ref{eq:intraBCD}, $\Omega_{vc}$ is the Berry curvature calculated for the insulating system that can be represented by a two-level system with a valence band ($v$) maximum and a conduction band ($c$) minimum (Fig.~\ref{fig:okyay2022second}(b)). Further, $\omega$ and $\Theta(2 \omega - \omega_{cv})$ represent the frequency of the sub-bandgap driving field and the Heaviside step function, respectively. As a result of the Heaviside step function, interband BCD is invariably zero below the incident light frequency, which equals half of the band gap. The second harmonic generation (SHG) susceptibility tensor $\chi^{(2)}$ has two components $\chi^{(2)}_{2-\textup{band}}$ and $\chi^{(2)}_{3-\textup{band}}$, given by

\begin{align}
    \chi^{(2)}_{2-\textup{band}} & = \frac{8 \pi e^{3}}{\hbar^{2}} \int_{k} \mathrm{Im} \sum_{ij} \frac{f_{ik}(1-f_{jk})v^{y}_{ij}v^{x}_{ji}}{\omega_{ji}^{4}}  (v_{jj}^{x} - v_{ii}^{x}) \:\delta(2\omega - \omega_{ji}),
        \nonumber \\
    \chi^{(2)}_{3-\textup{band}} & = - \frac{8 \pi e^{3}}{\hbar^{2}} \int_{k} \mathrm{Im} \sum_{ijl} \frac{f_{ik}(1-f_{jk})v^{y}_{ij}v^{x}_{ji}v^{x}_{li}}{\omega_{ji}^{3}(\omega_{il}+\omega_{il})}   \:\delta(2\omega - \omega_{ji}).
    \label{eq:2band}
\end{align}

Here $\int_{k} = \frac{1}{4\pi^2}\int d^{2}\vec{k}$ is the momentum space integration for two-dimensional systems, $v^{x/y}$ represents the band velocity and the difference $(v_{jj}^{x} - v_{ii}^{x})$ is taken at a crystal momentum $\vec{k}$. In the case when the contribution from the valence band top and conduction band bottom dominate the effects of the other bands, we can rewrite the Eq.~\ref{eq:2band} as 

\begin{equation}
    \chi^{(2)}_{2-\textup{band}} = \frac{\pi e^3}{\hbar^2 \omega^2} \sum_{ij} \vec{D}_{ij} \cdot \hat{x} \approx \frac{\pi e^3}{\hbar^2 \omega^2} (\vec{D}_{vc} \cdot \hat{x}).
    \label{eq:2bandandBCD}
\end{equation}

We note that $\chi^{(2)}_{2-\textup{band}}$ acts as a pseudovector and is odd under an improper rotation, i.e., mirror reflection. However, it is evident that there is no direct pseudovector form of the 3-band component of the SHG susceptibility, $\chi^{(2)}_{2-\textup{band}}$. To understand the symmetry influence on the 2-band and 3-band components, these parameters are calculated for crystals with different symmetry as shown in Fig.~\ref{fig:okyay2022second}(c). As an illustration, we consider two distinct crystal phases, i.e., 2H and 1T$_{d}$ phases of WTe$_{2}$ with crystal symmetry $D_{3h}$ and $C_{s}$, respectively~\cite{garcia2020canted,bastos2019ab}. The band gap of the 1T$_{d}$ phase ($\sim 0.14$ eV) is significantly smaller compared to that of the 2H phase ($\sim 0.63$ eV). As a consequence, the SHG susceptibility of 1T$_{d}$ phase is almost two orders of magnitude higher than that of the 2H phase, as presented in Fig.~\ref{fig:okyay2022second}(d). It is interesting to note that the 2-band response is precisely in line with the interband BCD contribution $D_{vc}/\omega^2$, which is a direct consequence of the relation given in Eq.~\ref{eq:2bandandBCD}. A critical observation also reveals that the 2-band contribution and hence the interband BCD is invariably zero for the 2H phase, while it is predominant in the 1T$_{d}$ phase. To explain this, we realize that although the magnitude of interband BCD corresponds to the system's band gap, its non-zero value is protected by the crystal symmetry of 1T$_{d}$ phase. In particular, similar to the conventional NLH effect, the interband BCD generated second harmonic Hall response strongly depends on the crystal symmetry, as we shall critically discuss in an upcoming sections.


\subsection{Disorder contributions to non-linear Hall response}

Before that, it is important to briefly discuss the other possible contributions to the NLH signal. We know that the interaction of Bloch electrons with disorder is notably influenced by their quantum geometry, which becomes especially pronounced in phenomena such as the linear anomalous Hall effect~\cite{nagaosa2010anomalous}. In particular, the nonzero Berry connection and curvature may arise due to asymmetric scattering processes, known as skew scattering and side jump~\cite{sinitsyn2006coordinate}. A comprehensive Boltzmann transport formalism for the conventional or linear anomalous Hall effect has been well-established~\cite{sinitsyn2007anomalous,sinitsyn2007semiclassical}. Nevertheless, recent efforts have extended the framework phenomenologically to comprehend second-order nonlinear responses at low frequencies~\cite{konig2019gyrotropic,isobe2020high,du2019disorder,morimoto2016semiclassical,nandy2019symmetry}. The framework has been further modified by introducing a new side-jump contribution that does not have any semiclassical correspondence~\cite{xiao2019theory}. Moreover, Du \textit{et al.} have formulated a purely quantum theory for the NLH effect in the presence of scattering employing a diagrammatic technique~\cite{du2021quantum}. Upon incorporating disorder contributions through Feynman diagrams, it has been found that the overall NLH conductivity gets amplified due to disorder.

A prolonged debate, spanning over a century, regarding the origin of the anomalous Hall effect~\cite{chang2023colloquium}, has reached a culmination by categorizing the mechanisms into intrinsic (disorder-free) and extrinsic (disorder-induced) contributions. In the context of NLH effects [Fig.~\ref{fig:intrinsicextrinsic}], disorder assumes a pivotal role compared to its linear counterpart~\cite{tian2009proper,hou2015multivariable}. This is primarily due to the inescapable influence of disorder scattering on the Fermi surface in the nonlinear case. To understand the effect of both intrinsic and extrinsic parts, let us consider a minimal low-energy model of a two-dimensional tilted Dirac cone, which reads

\begin{equation}
   H = t k_x \mathbb{I} + \nu k_x \sigma_x + \nu k_y \sigma_y + \Delta \sigma_z,
\label{eq:tilteddisorder}
\end{equation}

The above Eq.~\ref{eq:tilteddisorder} is identical to the one given in Eq.~\ref{eq:tightbindingthirdorder} under the assumption of isotropic velocity i.e. $v_x = v_y = \nu$. Here $\{\sigma_{x/y/z}\}$ is the triad of Pauli matrices, with $\mathbb{I}$ being the identity. The variable $2 \Delta$ represents the band gap of the massive Dirac model with tilt parameter $t$~\cite{du2018band}. Notably, looking over the time-reversal symmetry in a single Dirac cone in two dimensions is impracticable. To satisfy time-reversal symmetry, it is essential to include its time-reversed partner ($\Delta \rightarrow - \Delta$, $t \rightarrow - t$) in opposite regions of the Brillouin zone ($\vec{k} \rightarrow - \vec{k}$). Next, disorder can be introduced to the Hamiltonian, given in Eq.~\ref{eq:tilteddisorder}, as a spin-independent random potential $\mathcal{V} = \sum_{i} V_{i} \delta(\vec{r}-\vec{R}_{i})$, for both types of correlations -- Gaussian ($\langle V_{i}^{2}\rangle = V_{0}^2$) and non-Gaussian ($\langle V_{i}^{3} \rangle = V_{I}^{3}$). In the above case the intrinsic ($\chi^{in}_{yxx}$), side-jump ($\chi^{exsj}_{yxx}$), and skew-scattering ($\chi^{exsk1}_{yxx}$ and $\chi^{exsk2}_{yxx}$) response functions [illustrated in Fig.~\ref{fig:intrinsicextrinsic}(d)] can be found up to the linear order in $t$ as \cite{du2019disorder}

\begin{eqnarray}
    \chi^{in}_{yxx} &=& \frac{3 t \Delta \nu^2 e^3 (E_{F}^2 - \Delta^2)}{2 h n_i V_{0}^2 E_{F}^3 (E_{F}^2 + 3 \Delta^2)}, \nonumber \\
    \chi^{exsj}_{yxx} &=& \frac{t \Delta \nu^2 e^3 (E_{F}^2 - \Delta^2)  (E_{F}^2 - 25 \Delta^2)}{2 h n_i V_{0}^2 E_{F}^3 (E_{F}^2 + 3 \Delta^2)^2}, \nonumber \\
    \chi^{exsk1}_{yxx} &=& -\frac{t \Delta \nu^2 e^3 (E_{F}^2 - \Delta^2)^2  (13E_{F}^2 + 77\Delta^2)}{4 h n_i V_{0}^2 E_{F}^3 (E_{F}^2 + 3\Delta^2)^3}, \nonumber \\
    \chi^{exsk2}_{yxx} &=& -\frac{t\Delta \nu^2 e^3 V_{I}^{3}(E_{F}^2 - \Delta^2)^2  (5E_{F}^2 + 9\Delta^2)}{h n_i V_{0}^6 E_{F}^2 (E_{F}^2 + 3\Delta^2)^3}.
\label{eq:disorderexpression}
\end{eqnarray}

We note that the presence of a single mirror symmetry operation $k_y \leftrightarrow -k_y$ assigns a zero value of the component $\chi_{xyy}$. Furthermore, the direct current NLH signal exhibits angular dependence characterized by a one-fold pattern~\cite{kang2019nonlinear,du2018band}. From the above expressions given in Eq.~\ref{eq:disorderexpression}, it is clear that the skew-scattering predominantly depends on the non-Gaussian scattering strength $V_{I}$. Further, the nonlinear conductivity for all the cases depends on the parameter $E_{F}^{2}-\Delta^2$, ensuring the vanishing contribution near the band edge. Du \textit{et al.} have formulated symmetry analysis, which indicates the non-zero components of NLH response tensor in the case of two-dimensional point groups, which are provided in Fig.~\ref{fig:intrinsicextrinsic}(e)~\cite{du2021quantum}. The differentiation between various contributions of the anomalous Hall effect essentially relies on the distinctive scaling laws governing the relationship between the transverse Hall signal and the longitudinal signal~\cite{tian2009proper}. Similarly, a general scaling law for the NLH effect has also been proposed to separate out different intrinsic and extrinsic nonlinear contributions~\cite{du2021quantum}. In their recent study, Chen \textit{et al.} have explored the effect of Anderson disorder~\cite{lee1985disordered,evers2008anderson,RevModPhys.69.731,prodan2010entanglement,pozo2019quantization,garcia2015real,weisse2006kernel} on the NLH signal of real lattices~\cite{chen2023fluctuation}. It has been found that the introduction of disorder results in an increasing fluctuation of the NLH conductance as the Fermi energy transitions from the band edges to higher energy states~\cite{ma2019observation,ho2021hall}. This phenomenon is markedly distinct from the universal conductance fluctuations in the linear conductance regime, which are unaffected by the Fermi energy~\cite{lee1985universal,ren2006universal}. Consequently, the conventional disorder-free distribution of the Berry curvature cannot explain these intriguing fluctuations. Notably, a fascinating emergence of "Anderson localization" has been identified in the nonlinear response, which acts along the transverse direction and, in principle, differs significantly from the well-known linear response scenarios. In the following section, we shall focus on the detailed analysis of the symmetry-based indicators for the emergence of the NLH effects under time-reversal symmetric conditions.

As we have already discussed, the intrinsic BCD mechanism manifests nonlinear transport perpendicular to the current direction, i.e., the Hall response, while it does not exhibit such behavior along the longitudinal direction~\cite{ma2019observation,kang2019nonlinear}. On the contrary, extrinsic mechanisms stemming from skew scattering induce second-order transport in both transverse and longitudinal directions~\cite{du2019disorder,isobe2020high}. Consequently, examining the longitudinal nonlinear response aids in distinguishing between the origins of nonlinear transport. Notably, the occurrence of a finite second-order response from scattering does not necessitate the breaking of time-reversal symmetry -- it can emerge from a scattering process reflecting quantum wave functions through chirality~\cite{isobe2020high,tokura2018nonreciprocal,belinicher1980photogalvanic}. Furthermore, recent work reveals a substantial second-order electric transport in monolayer graphene–hBN van der Waals (vdW) heterostructures across both longitudinal and transverse directions, even without a magnetic field~\cite{he2022graphene}. This nonlinear conductivity, surpassing expectations by up to five orders of magnitude compared to single non-centrosymmetric crystals, is attributed to skew scattering driven by chiral Bloch electrons in graphene superlattices. These electrons derive from the valley-contrasting chirality inherent in gapped graphene, validated through direct measurements of second-order charge transport. Unlike optical second-harmonic generation (SHG) studies, measurements by He \textit{et al.} solely capture graphene's contributions, facilitating accurate evaluation of its nonlinear conductivity by excluding signals from the insulating hBN layer~\cite{he2022graphene}.

It is essential to differentiate between the diverse contributions to the nonlinear Hall signal. In the context of the anomalous Hall effect, this differentiation hinges upon the scaling law dictating the relationship between the transverse Hall response and the longitudinal signal~\cite{nagaosa2010anomalous}. 
In addressing the scaling law associated with the nonlinear Hall effect, typically a ratio is introduced: $V_{y}^{N}/(V_{x}^{L})^2 = \zeta_{yxx} \rho_{xx}$. Here $V_{y}^{N}$ represents the nonlinear Hall effect, while $V_{x}^{L}$ denotes the linear longitudinal voltage. The driving electric current is applied along the $x$ direction, while the measurement of the nonlinear Hall voltage is conducted along the $y$ direction~\cite{du2018band,du2019disorder}. 
One advantage of this variable is its ability to render the intrinsic and side-jump components independent of disorder. The general scaling law of the nonlinear Hall effect can be derived as

\begin{equation}
    \frac{{V_y^N}}{{(V_x^L)^2}} = {\cal{C}}^{in} + \mathop {\sum}\limits_i {\cal{C}}_i^{sj}\frac{{\rho _i}}{{\rho _{xx}}} + \mathop {\sum}\limits_{ij} {\cal{C}}_{ij}^{sk,1}\frac{{\rho _i\rho _j}}{{\rho _{xx}^2}} + \mathop {\sum}\limits_{i \in S} {\cal{C}}_i^{sk,2}\frac{{\rho _i}}{{\rho _{xx}^2}}.
    \label{eq:generaleqn}
\end{equation}

Here the coefficients representing the intrinsic, side-jump, intrinsic skew-scattering, and extrinsic skew-scattering contributions are denoted as $\mathcal{C}^{in}$, $\mathcal{C}{i}^{sj}$, $\mathcal{C}{i}^{sk,1}$, and $\mathcal{C}_{i}^{sk,2}$, respectively. The symbol $S$ represents static disorder scattering sources. Considering the major scattering sources, i.e., static ($i=0$) and dynamic ($i=1$), one can write the scaling law as 

\begin{equation}
    \frac{{V_y^N}}{{(V_x^L)^2}} = \frac{1}{{\rho _{xx}^2}}({\cal{C}}_1\rho _{xx0} + {\cal{C}}_2\rho _{xx0}^2 + {\cal{C}}_3\rho _{xx0}\rho _{xxT} + {\cal{C}}_4\rho _{xxT}^2),
\end{equation}

where the scaling parameters can be written as

\begin{equation}
\begin{array}{*{20}{l}} 
{{\cal{C}}_1 = {\cal{C}}^{sk,2},\,{\cal{C}}_2 = {\cal{C}}^{in} + {\cal{C}}_0^{sj} + {\cal{C}}_{00}^{sk,1},} \\ 
{{\cal{C}}_3 = 2{\cal{C}}^{in} + {\cal{C}}_0^{sj} + {\cal{C}}_1^{sj} + {\cal{C}}_{01}^{sk,1},} \\ 
{{\cal{C}}_4 = {\cal{C}}^{in} + {\cal{C}}_1^{sj} + {\cal{C}}_{11}^{sk,1}.} 
\end{array}
\end{equation}

The parameters outlined can be inferred from experimental data~\cite{tian2009proper,hou2015multivariable}. The residual resistivity due to static impurities at zero temperature and dynamic disorder at finite temperature are denoted by $\rho_{xxo}$ and $\rho_{xxT}$, respectively. In the limit of zero temperature, the scaling law adopts linearity, as dictated by the relationship $V_{y}^{N}/(V_{x}^{L})^2 \approx \mathcal{C}_1 \: \sigma_{xx0} + \mathcal{C}_{2}$.

By fitting the experimental data using this relation, the extrinsic skew-scattering coefficient $\cal{C}^{sk,2}$, can be empirically determined from the overall nonlinear Hall conductivity. Moreover, at finite temperatures, it is more convenient to reframe the scaling law as

\begin{equation}
    \begin{array}{*{20}{l}} {\frac{{V_y^N}}{{(V_x^L)^2}} - {\cal{C}}_1\sigma _{xx0}^{ - 1}\sigma _{xx}^2} \hfill & \simeq \hfill & {({\cal{C}}_2 + {\cal{C}}_4 - {\cal{C}}_3)\sigma _{xx0}^{ - 2}\sigma _{xx}^2} \hfill \\ {} \hfill & {} \hfill & { + ({\cal{C}}_3 - 2{\cal{C}}_4)\sigma _{xx0}^{ - 1}\sigma _{xx} + {\cal{C}}_4.} \hfill \end{array}
\label{eq:scaling2}
\end{equation}

In the above case (Eq.~\ref{eq:scaling2}), the scaling variable transforms into a parabolic function of $\sigma _{xx0}^{ - 1}\sigma _{xx}$, namely, $V_y^N/(V_x^L)^2 - {\cal{C}}_1\sigma _{xx0}^{ - 1}\sigma _{xx}^2$. By fitting the experimental data with a parabolic function, one can deduce the remaining scaling parameters~\cite{du2019disorder}.

\section{Symmetry indicators}

In this section, we start our discussion by highlighting the very first symmetry restriction to the conductivity tensor given in Eq.~\ref{eq:conductivity}, namely Onsager's reciprocal theorem. In a seminal work, while dealing with the irreversible processes in non-equilibrium statistical physics, Onsager introduced the famous reciprocity theorem~\cite{onsager1931reciprocal}. This states that for any non-equilibrium system the fluxes ($\mathcal{J}$) can be expressed as $\mathcal{J}_{i} = \sum_{j} \mathcal{L}_{ij} \mathcal{F}_{j}$, where $\mathcal{F}_{j}$ is the force responsible for dragging the system out of equilibrium, such that the kinetic coefficients $\mathcal{L}$ form a symmetric tensor, i.e., $\mathcal{L}_{ij}$ = $\mathcal{L}_{ji}$. The reciprocity theorem emerges due to the time-reversal symmetry inherent in the microscopic dynamics of non-equilibrium systems. Consequently, the linear conductivity tensor becomes symmetric when the preservation of time-reversal symmetry is ensured. When a force (such as an electric field) aligns with the principal axes of the system, its first-order response -- flux or current, in this case -- coincides directionally with the driving force. This leads to the inevitable disappearance of the conventional Hall effect, i.e., the anti-symmetric component of the conductivity tensor, when time-reversal symmetry is maintained. Therefore, one typically needs to introduce a time-reversal symmetry-breaking field for the linear Hall effect to occur, albeit the crystal symmetries should also be maintained~\cite{vsmejkal2020crystal}. We note that the above statement is only valid for the charge current; the spin Hall effect is still possible in time-reversal symmetric systems~\cite{sinova2004universal,kane2005quantum,bernevig2006quantum,oh2013complete,sinova2015spin}. 

The transformation condition for the charge Hall conductivity (linear response) can be expressed in terms of the crystal's point group symmetry ($\hat{P}$) as $\sigma_{\mu \nu}  = \mathrm{det}(P) \: \hat{P}\sigma_{\mu \nu} \hat{P}^{T}$. In two-dimensions, Hall conductivity acts as a pseudoscalar, $ \sigma_{Hall} = \epsilon_{\mu \nu} \frac{\sigma_{\mu \nu}}{2} \rightarrow \mathrm{det}(\hat{P}) \: \sigma_{Hall} $, where $\epsilon_{\mu \nu}$ is the Levi-Civita symbol. We note that the matrix $\hat{P}$ is, by definition, an orthogonal matrix, so that $\mathrm{det}(\hat{P}) = \pm 1$ condition always holds. However, the above discussion suggests that systems characterized by point group symmetry meeting the condition $\mathrm{det}(\hat{P}) = -1$ will not exhibit a linear Hall response.

In contrast, NLH conductivity is not subject to such constraints and can be observed even under time-reversal symmetric conditions. Nonetheless, in a time-reversal invariant system, the Berry curvature and its first-order moment, BCD, emerge only when the inversion symmetry is explicitly broken. Furthermore, the point group symmetry of any two-dimensional crystal satisfying the condition $\mathrm{det}(\hat{P}) = - 1$, entails a single mirror plane orthogonal to the system. Unlike linear response, the presence of a mirror plane in the system does not ensure the absence of a NLH signal; it only mandates the Hall current to flow in a direction perpendicular to it~\cite{sodemann2015quantum}. In other words, the mirror plane is the maximum symmetry allowed in any two-dimensional crystal to exhibit a NLH response -- BCD, a pseudovector in two dimensions, orients itself in a direction orthogonal to the mirror symmetry plane (see, Eq.~\ref{eq:nheexpress}). In this regard, Tsirkin \textit{et al.} have introduced a general prescription for separating the higher-order currents into Ohmic- and Hall-type responses in a crystal without considering its symmetry~\cite{tsirkin2022separation}. From the above discussion, it is evident that only a handful of crystallographic symmetries, for example, $C_{v}$, are allowed to exhibit the NLH response in the presence of inversion symmetry breaking potentials. Here, it is worth mentioning that, as we discussed, the Berry curvature undergoes complete cancellation across all momenta within the three-dimensional Brillouin zone for materials characterized by both time-reversal invariance and inversion symmetry. Consequently, such materials might appear less interesting, given that their anomalous electron velocities seemingly vanish. However, recent investigation challenges this notion by demonstrating that while this conclusion is valid for the bulk of three-dimensional materials, the opposite holds true at their surfaces and interfaces~\cite{wawrzik2023surface}. In particular, Bloch electrons exhibit a finite anomalous velocity on general Miller index surfaces, even in materials with bulk inversion and time-reversal symmetry. Interestingly, the surface Berry curvature becomes zero at unreconstructed low Miller index surfaces. On the contrary, any higher index surface, including those with surface steps that disrupt C$_2$-rotation around the surface normal (e.g., miscut and vicinal surfaces), is associated with a non-zero surface Berry curvature. Moreover, a finite surface Berry curvature dipole emerges when there is no more than one mirror plane in two dimensions. We note that this symmetry requirement is satisfied by a wide range of elementary surfaces, including in elemental Bismuth \cite{makushko2023tunable}. Another important class of material that will show significant surface BCD is the one containing Weyl Fermi arcs. Weyl Fermi arcs are typically associated with a surface Berry curvature divergence at a hot surface line, leading to a giant surface BCD~\cite{wawrzik2021infinite}. These avenues for generating BCD and resulting NLH effect remain to be explored in experiments.

In three-dimensional crystals, the symmetry constraints on the NLH effect are more lenient. In this case, the NLH conductivity can be expressed as a second-rank tensor that consists of nine distinct components. The usual convention to represent this tensor is to write it in terms of symmetric and skew-symmetric parts that transform independently under the symmetry operations. The skew-symmetric part acts as a vector and is firmly related to the presence of a polar axis in a crystal. As a consequence, out of the total of 32 point groups, only $\{C_v, C_{nv}\}$, where $n \in \{1,2,3,4,6\}$ with a polar axis can exhibit a nonzero skew-symmetric part. In the presence of a time-dependent but spatially uniform electric field, $\mathcal{E} = \mathrm{Re}[\mathcal{E}_0 \: \exp(i \omega t)], \; \mathcal{E}_0 \in \mathbb{C}$, the current $\vec{j} \: (= \vec{j}^0 + \vec{j}^{2 \omega}$) obtained from the skew-symmetric part can be expressed as follows,

\begin{align}
\vec{j}^0 &= \frac{e^3 \tau}{1+2 (i \omega \tau)} \left[ \vec{\mathcal{E}}^{*} \times (\vec{d} \times \vec{\mathcal{E}}) \right], \nonumber \\
\vec{j}^{2 \omega} &= \frac{e^3 \tau}{1+2 (i \omega \tau)} \left[ \vec{\mathcal{E}} \times (\vec{d} \times \vec{\mathcal{E}}) \right]. 
\label{eq:twocontributions}
\end{align}

Here the vector, $\vec{d}$, aligns itself in the direction of the polar axis of the crystal. In the above Eq.~\ref{eq:twocontributions}, we have obtained a purely rectified direct current, $\vec{j}^0$, in addition to a second harmonic response, $\vec{j}^{2 \omega}$. On the other hand, the symmetric part of the NLH response is non-vanishing for point groups $\{O, T, C_1, C_n, D_n \}, \; n \in \{2,3,4,6\} $, i.e., for non-centrosymmetric crystals without left-handed symmetries [$\mathrm{det}(\hat{P}) = -1$]. Only left-handed systems with mirror symmetry $C_{1v}$, $C_{2v}$, and $S_4$ are allowed for existence of the symmetric part of the NLH conductivity. Therefore, we have established that crystal symmetry plays a prominent role in determining the appearance and the direction of the NLH response in both two and three dimensions. In the following section, we shall critically discuss how the symmetry-based indicators mentioned above lead to the theoretical predictions of candidate materials with non-vanishing BCD-induced NLH response. 


\section{Material candidates}\label{sec:materials}

\begin{figure*}[t]
    \centering
    \includegraphics[width=14cm]{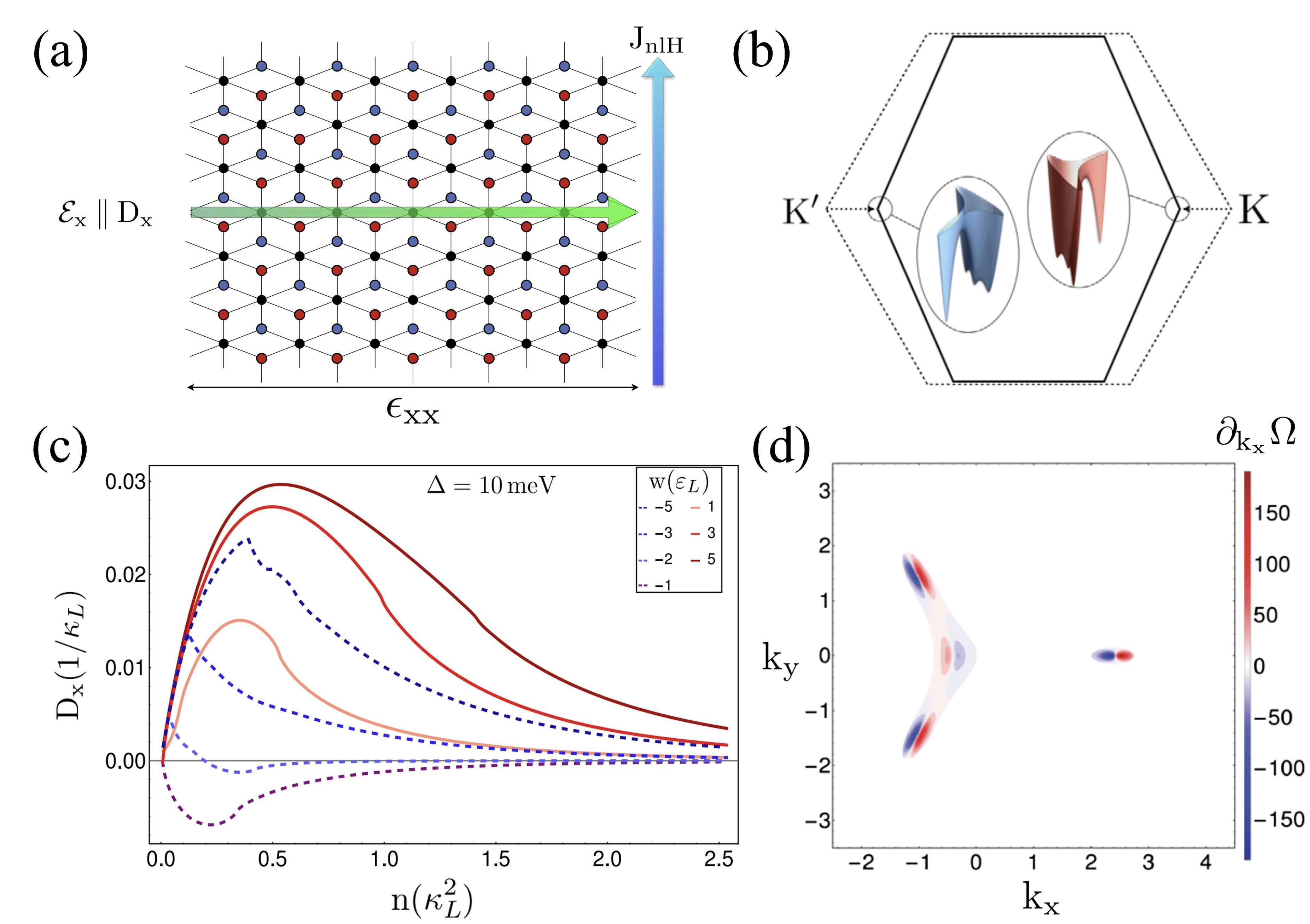}
    \caption{\textbf{BCD in strain-engineered graphene in presence of Fermi surface warping.} (a) The top view of bilayer graphene (Bernal-stacked) with broken inversion symmetry, achieved by maintaining different layers at different potentials indicated by blue and red colour spheres. Some of the top layer atoms sit exactly above the atoms of the bottom layer and are marked by black colour. For an in-plane electric field $\mathcal{E}_x$ applied parallel to the BCD direction, a NLH response ($J_{nlH}$) will be generated in a perpendicular direction. (b) The warped low-energy band structure for the strained bilayer graphene near two different valleys $K$ and $K^{\prime}$. (c) Variation of BCD with electron density for different values of uniaxial strain, $w$. The inversion symmetry breaking mass term $\Delta$ is fixed at 10 meV. (d) The gradient of the Berry curvature of the low energy conduction band along the $k_x$ direction. The strain value is $w = E_{L}$, where $E_{L}$ represents Lifshitz transition energy. Reproduced with permission from~\cite{battilomo2019berry}. }
    \label{fig:battilomo2019berry}
\end{figure*}

In this section, we shall first discuss the possibility of obtaining non-vanishing BCD in two-dimensional materials employing symmetry-based indicators presented above. Before that, it is worth exploring the potential origin of these geometric properties in the band structure utilizing some of the model Hamiltonians of well-established quantum materials. In recent years, there has been significant interest in the field of topological condensed matter physics, particularly in three-dimensional Dirac and Weyl semimetals. 
For example, a low-energy tilted massive Dirac type Hamiltonian that obeys all the symmetry restrictions for non-vanishing BCD is two-dimensions reads as
\begin{equation}
    H_{D} = t \: \tau_{z} \sigma_{0} k_y + (v_x k_x \sigma_y - v_y k_y \sigma_x) \otimes \tau_0 + \Delta \tau_{z} \sigma_{z}
    \label{eq:tilteddiracbcd}
\end{equation}
The above Eq.~\ref{eq:tilteddiracbcd}, is a general form of Eq.~\ref{eq:tightbindingthirdorder} but with a tilt along the $k_y$ direction. We note that here we have two distinct Pauli matrices, $\sigma$ and $\tau$ acting in the pseudospin space of the spin and valley degrees of freedom, respectively. 
As we have also mentioned earlier, the non-zero value of the tilt term $t$ is essential for non-vanishing BCD; the calculated Berry curvature~\cite{sodemann2015quantum,nandy2019symmetry} given below does not depend on it
\begin{equation}
    \Omega_{z} = \pm \frac{1}{2} \frac{\tau v_x v_y \Delta}{(\Delta^2+ v_x k_{x}^{2}+v_{y}k_{y}^{2})^{3/2}}.
\end{equation}
The $\pm$ sign represents the valence and conduction band and $\tau = \pm 1$ is for two distinct valleys. It is straightforward to show that only $D_{y}$ component of the BCD ($D$) will be nonzero in this case, which can be written as,
\begin{align}
    D_y = \sum_{\tau} \int \frac{d^2 k}{(2 \pi)^2} \Omega_{z} v_{k}^{y} \times \delta(\epsilon-\epsilon_{F}).
\end{align}
Here $\epsilon_{F}$ is the Fermi energy. The BCD value of each valley will be same and opposite in magnitude if we switch off the tilt term. Consequently the net BCD will be invariably zero. To avoid this situation we need a tilted Dirac model which provides the BCD value as
\begin{equation}
    D_y = \frac{3 t \Delta}{4 \pi \epsilon_{F}^{4}} [\epsilon_{F}^{2}-\Delta^{2}].
\end{equation}
The above expression is a simplified expression of BCD considering uniform velocity ($v_x = v_y$) and a small tilt parameter. Similarly in three dimension, Weyl semimetals have garnered attention for their ability to host gapless chiral quasiparticles, known as Weyl fermions, near the touching of a pair of nondegenerate bands, referred to as Weyl nodes~\cite{wan2011topological,burkov2011weyl,xu2011chern,xu2015discovery}. In these systems, nontrivial topological properties emerge due to Weyl nodes, which essentially serve as sources or sinks of Abelian Berry curvature. Consequently, three dimensional Weyl semimetals with tilted cones are extremely favorable platforms to evince strong nonlinear Hall response~\cite{matsyshyn2019nonlinear,singh2020engineering,zeng2021nonlinear}. A low-energy Hamiltonian that represents a pair of Weyl nodes under the influence of a mirror symmetry can be written as follows~\cite{facio2018strongly}

\begin{equation}
    H_{W} = v_x k_x \sigma_x + v_Z k_z \sigma_z + \frac{k_{y}^{2}-\lambda}{2 \Delta} \sigma_y + t k_x \mathbb{I},
    \label{eq:weylbcd}
\end{equation}

In the above Eq.~\ref{eq:weylbcd}, symbols have the same meaning as Eq.~\ref{eq:tilteddiracbcd}, and $\mathbb{I}$ represents the identity matrix. Given a finite carrier density ($n$) in the conduction band, and we are far from the critical point, BCD can be written as 

\begin{equation}
    d_z = \frac{(D_{xy} - D_{yx})}{2} \approx -\frac{n v_z t}{v_X \Delta}  \Delta_{m}^{-2}.
\end{equation}

Here $\Delta_m$ is the largest energy scale, and the tilt is minimal. Therefore, we have seen that NLH effect can be obtained in Weyl semimetals with broken inversion symmetry and tilted Weyl cones, and the response depends on the tilt parameter. Furthermore, the influence of surface states on the NLH response driven by the BCD in non-centrosymmetric, time-reversal invariant Weyl semimetals has been investigated~\cite{ovalle2022influence}. It has been revealed that the surface states, whether topological or not, play a more efficient role in contributing to the response compared to bulk states. This observation is not limited to topological semimetals and is expected to apply to both topologically trivial non-centrosymmetric materials and heterostructures. In subsequent work, an alternative mechanism, namely chiral-anomaly-induced NLH effect, has been introduced where Hall conductivity is proportional to $\vec{E} \cdot \vec{B}$~\cite{li2021nonlinear}. In particular, the tilting of Weyl cones, resulting from skewed energy dispersions around pairs of Weyl nodes, induces a NLH effect attributed to the collective actions of the anomalous velocity and the chiral anomaly. The above formalism has further been extended to the multi-Weyl semimetallic case~\cite{nandy2021chiral}. Furthermore, using the lattice Weyl Hamiltonian with an intrinsic chiral chemical potential, it has been discovered that planar electrical and thermoelectric
transport phenomena included nonlinear planar Nernst effect and nonlinear planar thermal Hall effect, induced by the chiral anomaly, can occur driven by electromagnetic gauge fields~\cite{zeng2022chiral}. Significantly, this holds true even for a pair of Weyl cones that are oppositely tilted or not tilted at all in the context of both time reversal and inversion broken Weyl semimetals.

Now, we will discuss how the physics mentioned above can be utilized to engineer the sizable NLH effect in real materials. To start with, the prototypical two-dimensional material graphene (point group $D_{6h}$) exhibits the maximum symmetry elements among all others in this family~\cite{geim2007rise,geim2009graphene,neto2009electronic}. The point group $D_{6h}$ of graphene can be easily recognized from the presence of a six-fold rotation, two-fold rotations, mirror symmetry planes, and the inversion symmetry. The discussion in the previous section shows that the symmetry elements of graphene must be significantly reduced to a single mirror line to obtain any finite BCD. The straightforward method to break the inversion symmetry in graphene is to place it on the top of a lattice-matched hexagonal boron nitride (h-BN) substrate~\cite{battilomo2019berry}. The h-BN structure offers an inversion symmetry breaking Semenoff mass~\cite{semenoff1984condensed} that results in an inversion symmetry breaking potential difference between the neighbouring sublattices. The Semenoff mass not only reduces the crystal symmetry to $C_{3v}$ but also splits the Dirac cones -- these can exhibit nonzero Berry curvature around the $K$ and $K^\prime$ points of the Brillouin zone. Furthermore, the application of an external strain along a crystallographic direction along the zigzag edge, represented by $\epsilon_{xx}$ in Fig.~\ref{fig:battilomo2019berry}(a), eliminates all the rotational symmetry elements and leaves alone the $M_x$ mirror line. Further, the Brillouin zone of graphene shrinks in the direction of applied uniaxial strain and the Dirac cones also shifts from the high symmetry points as depicted in Fig.~\ref{fig:battilomo2019berry}(b). The appearance of BCD perpendicular to the mirror plane is now allowed from the symmetry considerations. One can understand its emergence using a low-energy model by incorporating the anisotropic Fermi velocity due to strain-momentum coupling~\cite{guinea2010energy,de2012space} and an extra tilt term caused by an in-plane electric field~\cite{sodemann2015quantum}. However, an alternative framework has been proposed by Battilomo \textit{et al.} by considering the Fermi surface warping effect apart from the above-mentioned tilt mechanism~\cite{battilomo2019berry}. In the presence of strain-induced anisotropic velocities, it turns out that considering an isotropic trigonal ($C_3$ symmetry) warping is sufficient to generate non-vanishing BCD in graphene.  Nevertheless, the general model Hamiltonian with nonuniform warping term can be written as 

\begin{equation}
    H_{\mathrm{monolayer}} = \xi v_x k_x \sigma_x + v_y k_y \sigma_y + \frac{\Delta}{2} \sigma_{z} + (\lambda_1 k_{y}^{2} - \lambda_2 k_{x}^{2}) \sigma_x + 2 \xi \lambda_3 k_x k_y \sigma_y.
    \label{eq:graphenewarpingmono}
\end{equation}

Here $v_x$ and $v_y$ are the anisotropic velocities along $x$ and $y$ directions, respectively, the \{$\lambda_1, \lambda_2, \lambda_3$\} are the warping terms, $\xi = \pm 1$ is the valley index, and $\Delta$ represents the Semenoff mass. However, the Bernal-stacked bilayer graphene~\cite{mccann2013electronic} has an additional advantage over the monolayer case as the inversion symmetry can be broken by employing perpendicular electric fields only~\cite{yoon2017broken}. In the case of inversion symmetry broken bilayer graphene, the low energy Hamiltonian is modified as 

\begin{equation}
    H_{\mathrm{bilayer}} = \left[ \xi v_{\perp} k_x - \frac{(k_{x}^{2} - k_{y}^{2})}{2m} + w  \right] \sigma_x - \left[ \xi v_{\perp} + \frac{k_x k_y}{m} \right] \sigma_y + \frac{\Delta}{2} \sigma_z.
    \label{eq:bilayerwarping}
\end{equation}

Here the Fermi velocity corresponding to the inter-layer skew hopping is represented by $v_{\perp}$, whereas $w$ stands for the strain term. The pristine bilayer graphene ($\Delta = w = 0$) exhibits a Lifshitz transition~\cite{mccann2006landau,lemonik2010spontaneous,de2022cascade} characterized by splitting of the Fermi surface around the energy value $E_L=\frac{m v_{\perp}^{2}}{2}$. The Lifshitz transition results in a central Dirac cone along with shifted ``three leg" Dirac cones, with distinct Berry phases around both $K$ and $K^{\prime}$ points, as shown in Fig.~\ref{fig:battilomo2019berry}(b). The shift of the Dirac cones due to the Lifshitz transition can be measured in terms of a characteristic momentum $k_L = m v_{\perp}/\hbar$. 
We note that an applied strain in the inversion symmetry broken bilayer graphene gives rise to a tunable BCD. In particular, both the magnitude and sign of BCD as a function of electron density can be altered using external strain when the electric field generated mass term is kept constant (say, $\Delta$ = 10 meV) as illustrated in Fig.~\ref{fig:battilomo2019berry}(c). The BCD density, which results in the BCD upon Brillouin zone integration, is symmetric in nature for the transformation $k_y \rightarrow -k_y$ [Fig.~\ref{fig:battilomo2019berry}(d)]. This above symmetry is a clear evidence of the presence of time-reversal symmetry and a mirror symmetry in the system that also follows the crystal symmetry. Therefore, an external strain significantly modifies BCD density distribution in the momentum space. This symmetry-reduced deformed BCD density gives rise to an appreciable BCD in bilayer graphene. Quantitatively, the value of BCD in bilayer graphene is orders of magnitude larger than that of the monolayer case. Furthermore, in a recent study, Kheirabadi \textit{et al.} have studied the occurrence of the NLH effect by applying a time-reversal symmetry breaking in-plane magnetic field in place of inversion symmetry breaking perpendicular electric field~\cite{kheirabadi2022quantum}. In the latter case, non-vanishing Berry curvature and BCD result from the time-reversal symmetry breaking, although the crystal symmetry restrictions will be similar. Similar to the NLH effect, the linear electro-optic response is primarily dominated by by the BCD. We note that in most of the two-dimensional systems, the BCD is typically finite only with substantial spin-orbit coupling or higher-order Fermi surface warping. However, it has been found that the merging \cite{pal2023nonlinear} of untilted and unwarped Dirac points (semi-Dirac node) can also lead to a non-zero BCD, which is independent of Dirac velocity~\cite{samal2021nonlinear}. In other words, a pair of closely situated Dirac nodes with a saddle point in between, or a Dirac system where two nodes merge at a single point, serves as a platform for achieving a substantial BCD-generated NLH effect. These systems naturally exhibit helicity-dependent photocurrents due to linear momentum-dependent Berry curvatures.

\begin{figure*}[t]
    \centering
    \includegraphics[width=16cm]{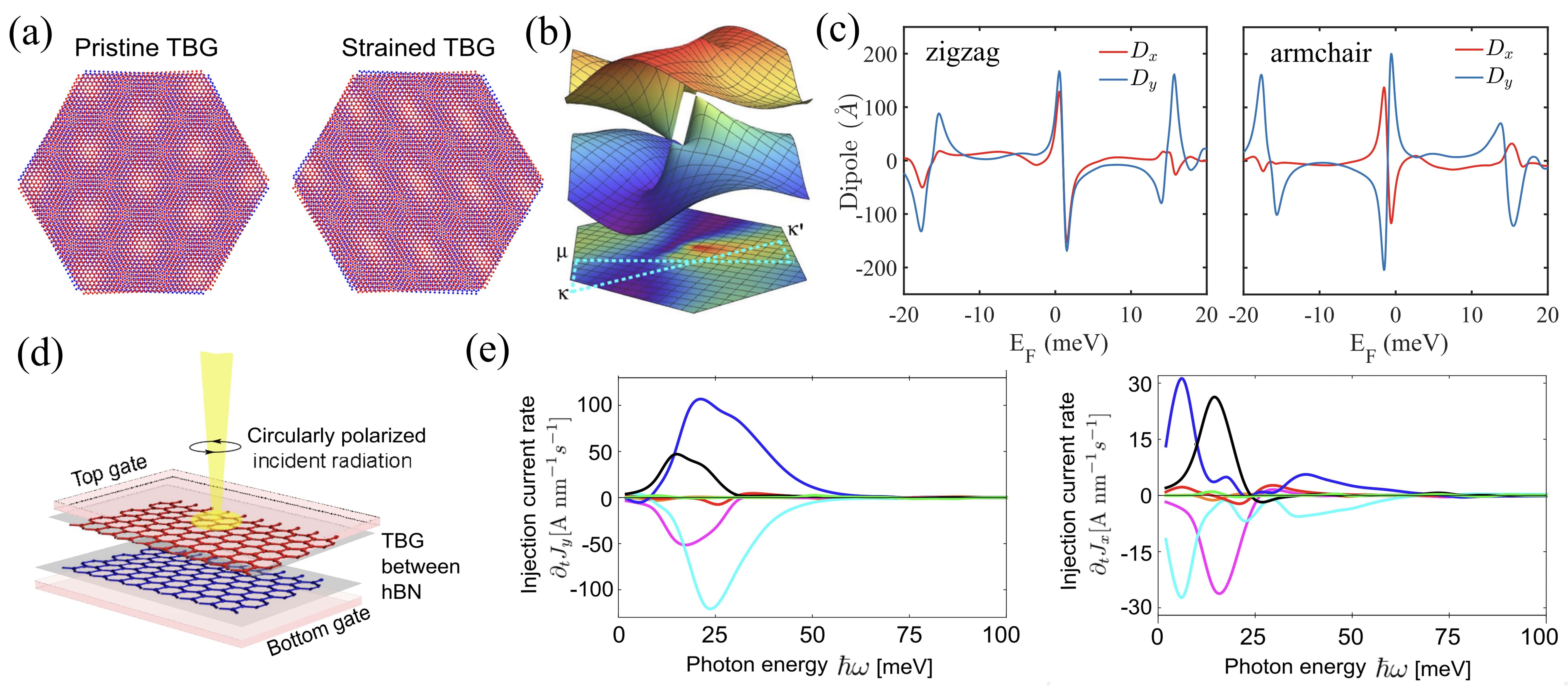}
    \caption{\textbf{Engineering Berry curvature dipole in encapsulated twisted bilayer graphene.} (a) Schematic of the top view of moir\'e superlattice for pristine (left panel) and strained (right panel) twisted bilayer graphene (TBG). Strained TBG exhibits an elliptical moir\'e pattern. (b) Distorted band structure of TBG in presence of strain. Unlike the pristine case, the degeneracy of the Dirac cones is broken in the strained case. The band projection on the moir\'e Brillouin zone is represented by the color scheme (maximum = red, minimum = purple) on the bottom hexagon. The symmetry path is indicated by a dotted line. (c) The BCD component $D_x$ and $D_y$ in TBG with twist angle $1.2^\circ$ in the presence of 0.1\% strain applied along zigzag (left panel) and armchair (right panel) direction. (d) Schematic representation of the device for obtaining circularly polarized photocurrent using strained TBG encapsulated with hexagonal boron nitride. (e) Injected current for circularly polarized excitation plotted for different chemical potential values. Here the bands near the charge neutrality point are primarily considered for the twist angle $1.05^\circ$. The orange, magenta, cyan, red, blue, black, and light green colours represent the chemical potential values $-50$ meV, $-10$ meV, $-2$ meV, $1.5$ meV, $5$ meV, $10$ meV, and $50$ meV, respectively. Reproduced with permission from~\cite{pantaleon2021tunable,zhang2022giant,arora2021strain}. }
    \label{fig:tbgbcd}
\end{figure*}
             
The twisted bilayer graphene (TBG) systems~\cite{bistritzer2011moire,tarnopolsky2019origin,yankowitz2019tuning,kerelsky2019maximized,andrei2020graphene,oh2021evidence,song2022magic,huang2022observation,diez2023symmetry} are another exciting platform to obtain significantly robust and highly tunable NLH responses. The pristine TBG possesses both inversion and time-reversal symmetry that does not allow any BCD in the system. However, the combined effect of the inversion symmetry breaking using h-BN substrate and strain engineering in the TBG forming moir\'e pattern [Fig.~\ref{fig:tbgbcd}(a)] helps to achieve the desired symmetry conditions for obtaining giant NLH effect at any filling factor~\cite{pantaleon2021tunable,zhang2022giant,arora2021strain}. In particular, applying opposite strain in the two distinct layers reduces the crystal symmetry of the TBG by breaking the $C_3$ axis, although inversion symmetry remains unaltered. We note that, under the influence of an external strain, the TBG systems manifest Dirac crossings away from the high symmetry points of the moir\'e Brillouin zone, and the corresponding energy values are also different [Fig.~\ref{fig:tbgbcd}(b)]. Furthermore, encapsulating TBG with h-BN invariably breaks the inversion symmetry of these systems and allows a finite Berry curvature to exist~\cite{cea2020band,pantaleon2021tunable}. In this case, the momentum space distribution of Berry curvature becomes highly nonuniform in nature, which invariably facilitates the existence of BCD. Near the magic angle (twist angle $\approx 1.2^\circ$) and a low uniaxial strain value ($\approx 0.1 \%$) along zigzag and armchair direction, one obtains two different BCD components, $D_x = \frac{\partial \Omega_z}{\partial x}$  and $D_y = \frac{\partial \Omega_z}{\partial y}$, where $x$ and $y$ are defined by the angular bisector between the two zigzag and armchair directions of the two layers, respectively. A strong BCD response is obtained for both strain types that can be easily externally tuned by a gate voltage, as represented in Fig.~\ref{fig:tbgbcd}(c). Notably, BCD changes its direction when the Fermi level crosses any band anticrossing point because of the sign change of Berry curvature~\cite{zhang2022giant}. Besides external bias, these topological transitions in the TBG system can be achieved via other functionalizations, such as doping. The above observation regarding the sensitivity of BCD on band inversions has been supported by a subsequent study~\cite{pantaleon2022interaction}, which also reveals the enhancement of BCD while incorporating long-range electron-electron Coulomb interaction. This observation is promising as specific fillings with an integer number of electrons/holes per moir\'e unit cell trigger strong electron-electron interactions leading to correlated phases in TBG~\cite{cao2018correlated,yankowitz2019tuning,xie2019spectroscopic}. In such cases, the conventional Fermi liquid model supporting the NLH transport theory introduced by Sodemann and Fu~\cite{sodemann2015quantum} is not expected to hold. Nevertheless, the observation of significant NLH effect in near-magic-angle TBG remains valid for the low filling factors that stay far from the special values $1/4$, $1/2$, and $3/4$~\cite{zhang2022giant}. In this case, an open problem will be exploring the effect of incorporating the Fock term in the Hamiltonian on the BCD. The Fock term is well recognized for inducing diverse symmetry-broken ground states, possibly influencing fillings distant from charge neutrality~\cite{cea2020band,pantaleon2022interaction}. 
In particular, BCD in strained (twisted) bilayer graphene gives rise to a "distributed transistor"-like response manifesting nonreciprocal optical interactions~\cite{rappoport2023engineering}. It has been reported that the optical amplification of incident light traversing the biased system is largely influenced by the polarization of the light.

Furthermore, the small band gap (of the order of a few meVs) at the charge neutrality point in the h-BN encapsulated stained TBG leads to a large injection current corresponding to the bulk photovoltaic effect~\cite{hunt2013massive}. The injection current in the inversion asymmetric systems is primarily a second-order DC response to an external alternating electric field, as depicted in Fig.~\ref{fig:tbgbcd}(d). As discussed in the previous section, this nonlinear photovoltaic current caused by the photoinduced transition between flat and remote bands can be well-explained by the interband BCD, as defined in Eq.~\ref{eq:intraBCD}. In particular, the interband BCD, here, has a maximum contribution near the $\Gamma$ point and systematically decreases by moving away from it. From the electronic band structure calculations it is evident that this interband BCD corresponds to the transition between the energy bands and lies within the energy window of a few meV. As a consequence, the variation of injection current rate with incident photon frequency exhibits noticeable peaks in the THz regime [Fig.~\ref{fig:tbgbcd}(e)]. 

\begin{figure*}[t]
    \centering
    \includegraphics[width=12cm]{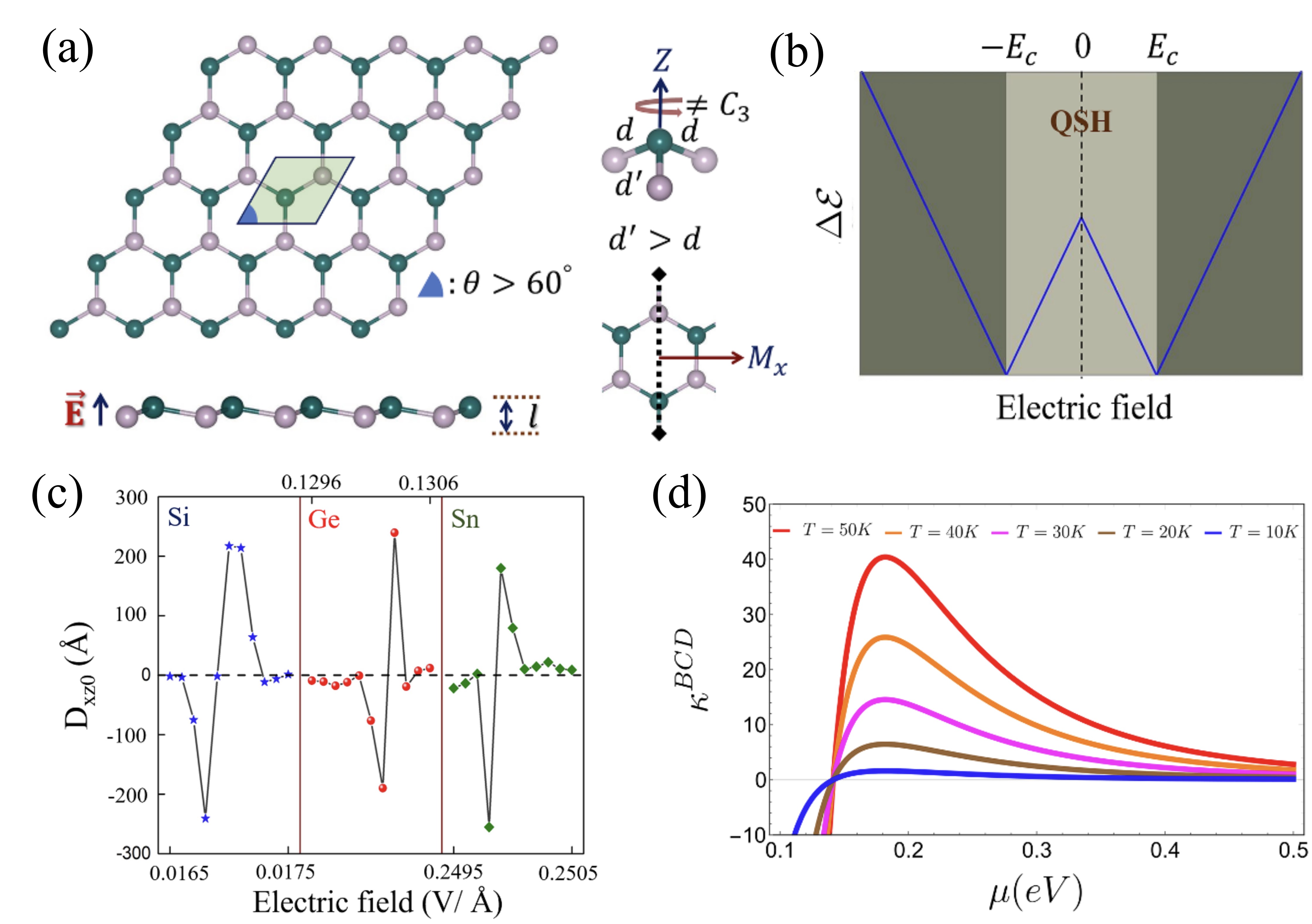}
    \caption{\textbf{Electric field tuned Berry curvature dipole in buckled honeycomb lattices.} (a) The schematic (top and side view) of the honeycomb lattices with a buckling $l$ and a perpendicular electric field $\vec{E}$. The electric field offers an inversion symmetry breaking Semenoff mass term generated by the potential difference between the neighbouring atoms (marked by different colours). The uniaxial strain engineering (2\%) modifies the lattice parameters and the angle between them. The only symmetry present in the structure is a single mirror line $M_x$. (b) The band gap variation ($\Delta \mathcal{E}$) of these systems with the external electric field. Finite spin-orbit coupling induces a topologically nontrivial band gap in the system that reduces up to a critical electric field $E_c$, where the systems behave as semimetals. About this critical value, the systems undergo a topological phase transition from the quantum spin Hall (QSH) state to the normal insulating state. (c) The BCD at the Fermi level $D_{xz0}$ shows a similar trend for silicene, germanene, and stanene, characterized by a maximum value that flips its direction about the topological quantum critical point. (d) The variation of the BCD generated thermal conductivity, $k^{BCD}$, with chemical potential, $\mu$, for different temperatures expressed in the unit $10^{-2} k_{B}^{2} / \hbar$ \AA{}. Reproduced with permission from \cite{bandyopadhyay2022electrically}. }
    \label{fig:our2dmat}
\end{figure*}

In comparison to planar structure graphene, the buckled honeycomb lattices, i.e., silicene, germanene, and stanene have some advantages to obtain sizable BCD~\cite{zhao2016rise,shan2023recent,acun2015germanene,balendhran2015elemental,ng20232d,rani2023stanene,lyu2019stanene}. For example, the above-mentioned buckled honeycomb systems exhibit appreciable spin-orbit coupling (SOC) values that induce a band gap in the system protected by a nontrivial topological index ($\mathbb{Z}_2 = 1$)~\cite{ezawa2015monolayer}. In other words, these systems host a quantum spin Hall (QSH) state featuring quantized Hall conductance without applying any magnetic field. However, an external electric field applied perpendicular to the system plane breaks the inversion symmetry of the system by introducing a Semenoff term in the model. The Semenoff mass invariably causes a topological phase transition from a QSH state to a normal insulating state beyond a system-specific critical value. Furthermore, the point group symmetry of these systems, $D_{3d}$, is lowered compared to graphene because of the absence of a six-fold rotational axis ($C_6$) and horizontal mirror plane ($\sigma_h$). Bandyopadhyay \textit{et al.} have proposed to reduce the symmetry of the buckled systems further down to a single mirror line by a combined effect of perpendicular electric field and uniaxial strain engineering as depicted in Fig.\ref{fig:our2dmat}(a)~\cite{bandyopadhyay2022electrically}. The strain value ($\approx 2\%$) is chosen so that it does not break the QSH state of the system. Therefore, the perpendicular electric field drives the strained buckled system through a topological phase transition [Fig.~\ref{fig:our2dmat}(b)] around the critical point $E_C \approx \lambda_{so}/l$, where $\lambda_{so}$ and $l$ represent spin-orbit coupling strength and buckling in the system, respectively. The mirror plane, $M_x$, of these systems indicates that the direction of the BCD obtained from the momentum integration of $d_{xz} = \frac{\partial \Omega_z}{\partial x}$ ($\Omega_z$ is the Berry curvature) will be along the $x$-axis. These systems exhibit a significantly large Berry curvature near the Fermi level, which results in a giant BCD ($\sim$ few nm) denoted by the maximum value $D_{xz0}$. The electric field efficiently controls the $D_{xz0}$ value, which undergoes a sharp sign change while going from the nontrivial phase to a normal insulating phase as shown in Fig.~\ref{fig:our2dmat}(c). The sign change of $D_{xz0}$ can be well understood from the flipping of Berry curvature of a particular band in a topological phase transition. Similar switchable BCD around the critical point of phase transition has also been observed in the case of two-dimensional class-AI metals~\cite{liao2021nonlinear}. Further, a mechanism is put forward to elucidate the generation of a second-order NLH effect in Kane-Mele type two-dimensional topological insulators~\cite{yang2011time}. 
In recent findings, a newly discovered group of materials called Jacutingaite has been revealed to exhibit a similar Kane-Mele type quantum spin Hall phase~\cite{marrazzo2018prediction}. Here breaking spatial and time reversal symmetry in the presence of Zeeman and Rashba couplings plays a crucial role in determining transport behaviour. Additionally, it has been established that the NLH effect exhibits distinctive signatures of topological phase transitions that can be achieved by tuning the energy gaps using external electromagnetic fields~\cite{malla2021emerging}. The two-dimensional topological insulating phase exhibits gapless helical edge states with quantized two-terminal conductance protected by time-reversal symmetry. However, electron-electron interactions and momentum-dependent spin polarization feedback under an applied current can open a gap in the edge state dispersion, breaking protection against elastic backscattering. Furthermore, this current-induced gap gives rise to a nonlinear contribution in the $I-V$ characteristic~\cite{balram2019current}. It has been further shown that BCD also allows an intrinsic contribution of the nonlinear thermal Hall effect in these buckled systems. As expected, BCD-induced nonlinear thermal Hall conductivity is proportional to the square of the temperature difference. Crystal symmetry implies that the $D_{xz}$ component of the BCD will be nonzero and its peak value can be controlled by changing the temperature of the system [Fig.~\ref{fig:our2dmat}(c)]. 

The lattice symmetry of the puckered honeycomb lattices such as black phosphorus (or phosphorene) are even lower compared to buckled lattices mentioned above~\cite{li2014black}. The lower symmetry of phosphorene can be understood from the point group $D_{2h}$, as the three-fold rotational symmetry is inherently missing. It has been shown that the inversion symmetry of phosphorene can be broken using a monochalcogenide substrate, for example, SnSe~\cite{muzaffar2021epitaxial}. This substrate effect is similar to applying a staggered onsite potential that not only breaks inversion symmetry but also reduces the crystal symmetry to a single mirror line~\cite{low2015topological}. Nevertheless, the wide band gap in the phosphorene structure does not allow non-vanishing BCD near the charge neutrality point. This problem can be solved by applying strain in the system. It has been proposed that an external strain can reduce the band gap significantly, which gives rise to an appreciable BCD in single- and multi-layered phosphorene~\cite{bandyopadhyay2023berry}.        


\begin{figure*}[h]
    \centering
    \includegraphics[width=14 cm]{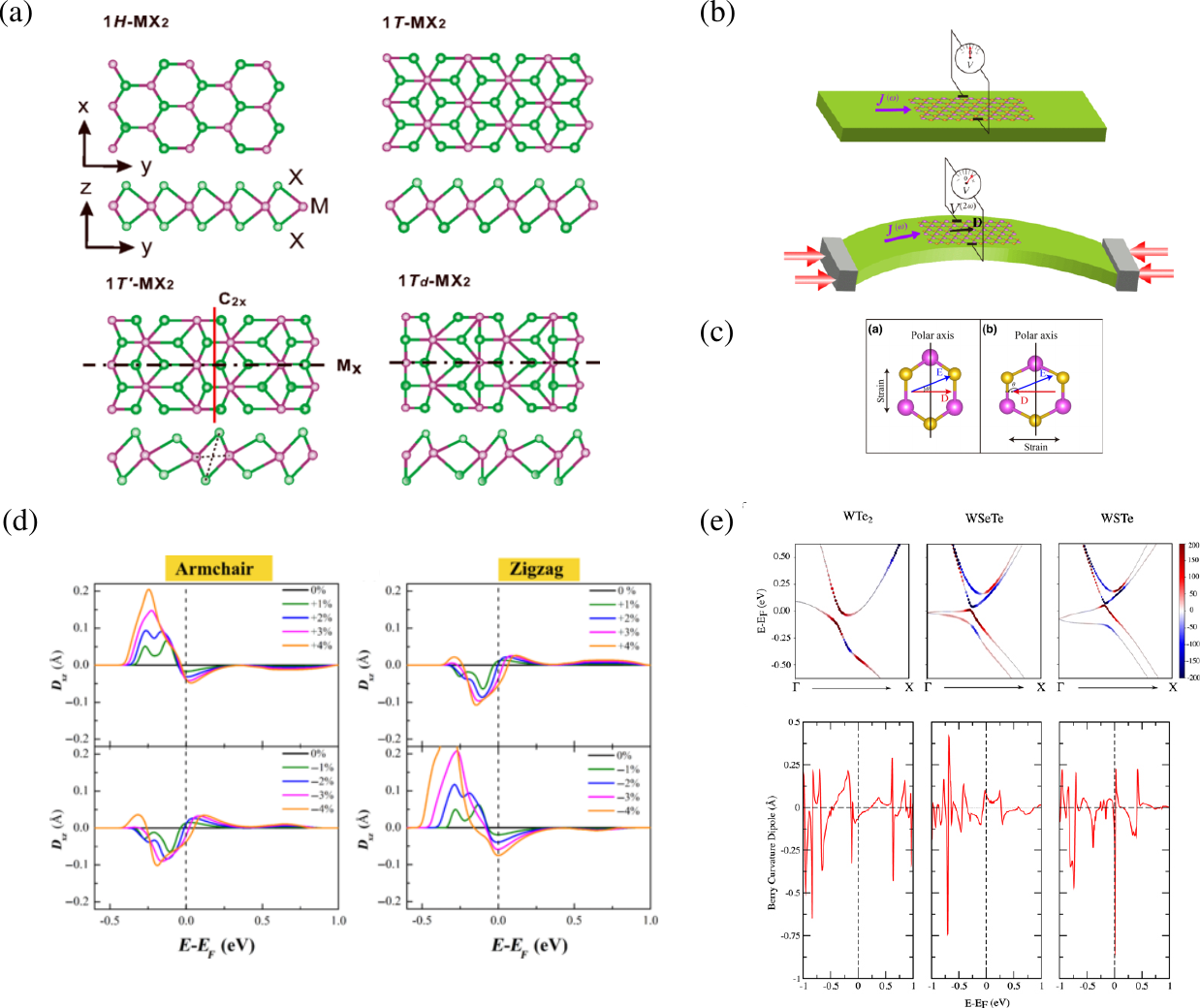}
    \caption{\textbf{Berry curvature dipole in transition metal dichalcogenides (TMDCs).} (a) Different phases associated with the TMDC family. A variety of methods can be used to induce and tune BCD and NLH effect in these materials. Reproduced with permission from~\cite{you2018berry}. (b) Schematic showing strain-induced piezoelectric-like effect in two-dimensional metallic 1H \ch{MX2} (M=Nb, Ta; X=S, Se) due to BCD. (c) Relationship between strain, BCD, andNLH current in the 1H phase of TMDCs. (d) Induced BCD in 1H \ch{NbS2} under uniaxial strain along two different directions. The positive (negative) values represent tensile (compressive) strain. Reproduced with permission from~\cite{xiao2020two}. (e) The distribution of Berry curvature and BCD associated with 1T' phase of \ch{WTe2} and the derived Janus monolayers WSeTe and WSTe. Reproduced with permission from~\cite{joseph2021tunable}. }
    \label{fig:theory_tmdc1}
\end{figure*}

\begin{figure*}
    \centering
    \includegraphics[width=12cm]{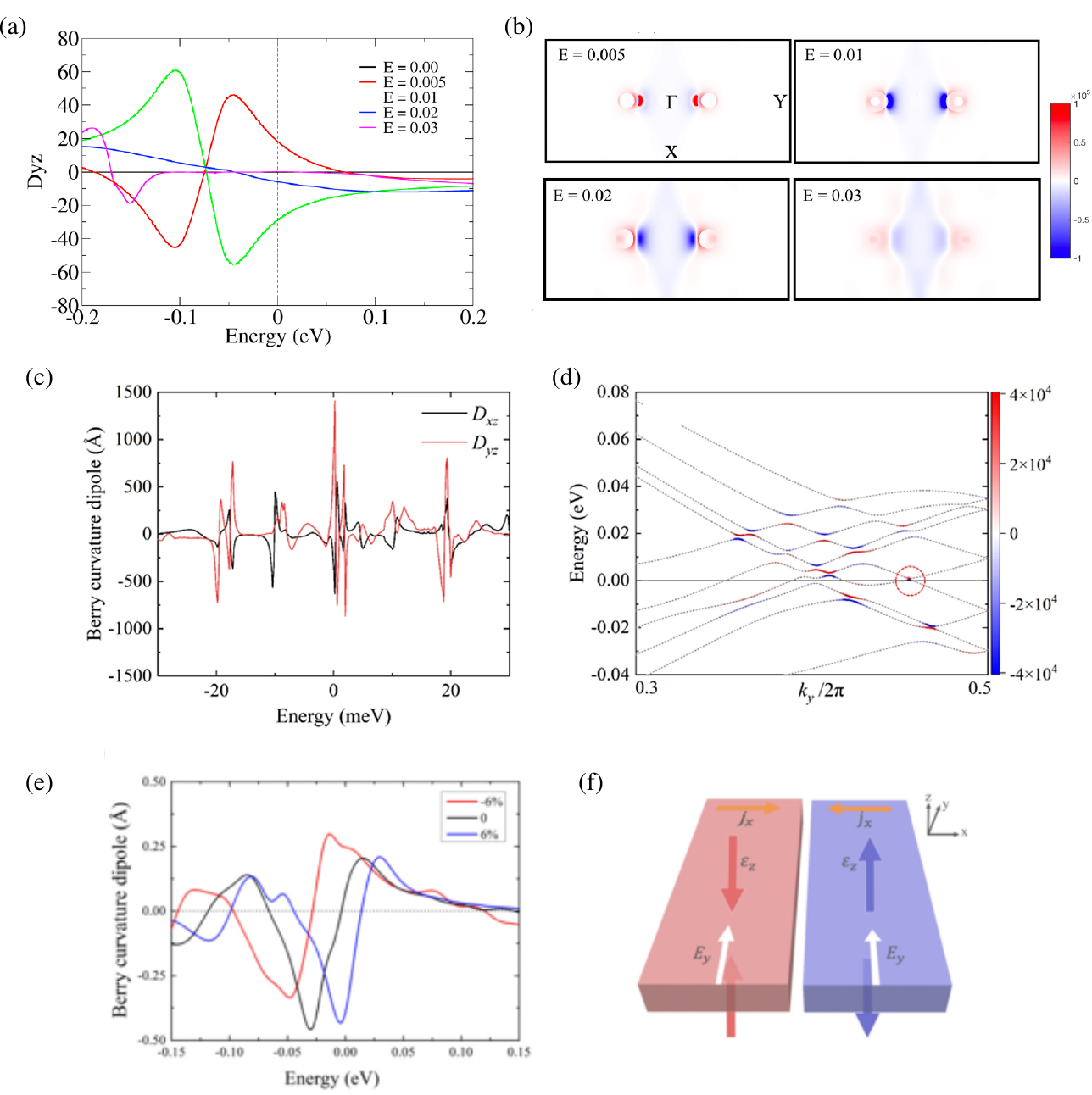}
    \caption{\textbf{Induced tunable Berry curvature dipole in \ch{WTe2} and TMDC heterostructures}. (a) The BCD and (b) its distribution in the two-dimensional Brillouin zone in 1T' monolayer of \ch{WTe2} with applied out-of-plane electric field. Reproduced with permission from~\cite{zhang2018electrically}. (c) Giant BCD and (d) the Berry curvature distribution along $\Gamma-Y$ direction in twisted bilayer \ch{WTe2}. The large value of BCD is due to extensive band crossings and band inversions present near the Fermi level. Reproduced with permission from~\cite{he2021giant}. (e) BCD under applied out-of-plane strain in \ch{MoSe2}/\ch{WSe2} vdW heterostructure. (f) Schematic of strain-gated NLH effect in \ch{MoSe2}/\ch{WSe2} heterostructure. The applied strain ($\epsilon_z$) can switch the direction of induced NLH current ($j_x$). Reproduced with permission from~\cite{jin2021strain}. }
    \label{fig:theory_wte2}
\end{figure*}

\begin{figure*}
    \centering
    \includegraphics[width=13cm]{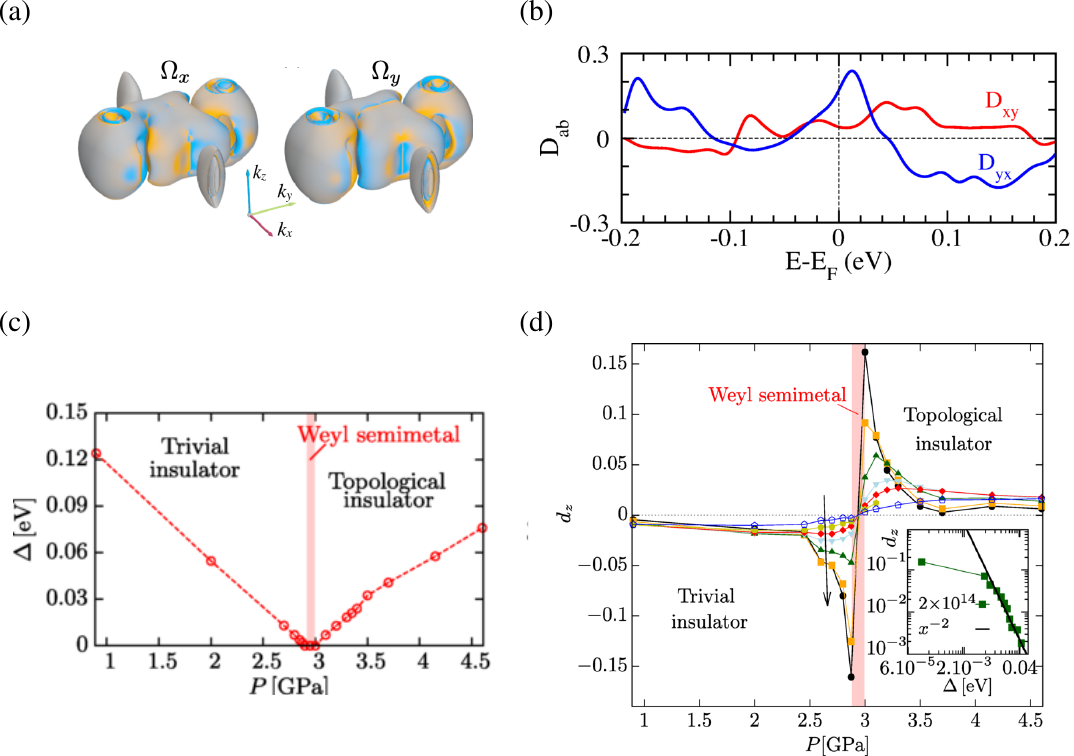}
    \caption{\textbf{Berry curvature dipole in bulk materials}. (a) Berry curvature on the Fermi surface and (b) non-zero BCD components of bulk T$_d$ \ch{MoTe2}. Yellow and blue represent positive and negative Berry curvature, respectively. Reproduced with permission from~\cite{singh2020engineering}. (c) Bandgap and (d) BCD of BiTeI as a function of pressure. The BCD peaks due to band gap closing and flips sign as a result of band inversion during the topological phase transition. Inset of (d) shows BCD as a function of bandgap. Reproduced with permission from~\cite{facio2018strongly}. }
    \label{fig:theory_bulk}
\end{figure*}

Transition metal dichalcogenides (TMDCs) are another promising class of materials that satisfy the necessary conditions to explore NLH effect.
TMDCs are among the original candidate materials proposed by Sodeman and Fu~\cite{sodemann2015quantum} to display NLH effect, and have been extensively studied because of the large spin-orbit coupling, low symmetry structures, and substantial Berry curvature.
This class of layered materials exists in three different metal-chalcogen coordination (phases) -- trigonal prismatic (H), octahedral (T) and distorted octahedral (T'). T$_d$, a variation of the T' phase without the $C_2$ rotation, has also been identified for some of these materials [Fig.~\ref{fig:theory_tmdc1} (a)].

The presence of large spin-orbit coupling and massive Dirac cone makes the 1H phase of two-dimensional TMDCs prospective candidates for a sizable BCD. 
The broken inversion symmetry in this trigonal prismatic coordinated system renders the Berry curvature finite, but the three-fold rotation symmetry forces the BCD to vanish.
Application of uniaxial strain breaks this symmetry, leading to a non-zero BCD, as demonstrated in the case of \ch{MoS2}~\cite{son2019strain} and \ch{WSe2}~\cite{you2018berry}. The application of strain reduces the symmetry, hence satisfying the additional symmetry constraint, leading to a non-zero NLH current in these materials. 
1H \ch{MX2} (M=Nb, Ta; X=S, Se) also displays such strain-induced BCD, leading to the possibility of piezoelectric-like behavior in these metallic systems~\cite{xiao2020two}. 
Similar to traditional piezoelectric materials based on the electric dipole, here, the absence of strain eliminates BCD, and hence, no NLH current is observed; on the application of strain, the additional symmetry is broken, leading to non-zero BCD and NLH current [Fig.~\ref{fig:theory_tmdc1} (b)-(d)]. 
NLH effect is also proposed to be an effective transport method to detect the polar order in ferroelectric-like metals \ch{LiOsO3}~\cite{xiao2020electrical} and SnTe~\cite{kim2019prediction,wang2019ferroicity}.
Zhou \textit{et. al.} also demonstrated that the construction of strained polar TMDCs, such as strained MoSSe, significantly enhances spin-orbit coupling in these systems, which leads to a stronger BCD~\cite{zhou2020highly}.

While the 1T phase of TMDCs has inversion, the distorted phases 1T' and T$_d$ do not, the latter more prominently than the former, thus leading to smaller values of BCD in 1T' structures.
Different approaches to largely break inversion symmetry to attain higher values of BCD have been proposed in these materials. 
One such approach, proposed by Joseph \textit{et al.}, is the construction of Janus monolayers of the 1T' phase~\cite{joseph2021tunable}. 
Janus monolayers are constructed by replacing one layer of chalcogen with its congeners, breaking the out-of-plane symmetry and, hence, the inversion. 
Replacing one layer of Te in 1T'~\ch{WTe2} with either Se or S (giving WSeTe or WSTe) significantly enhances the BCD as a result of the Berry curvature being largely concentrated near the reduced bandgap [Fig.~\ref{fig:theory_tmdc1} (e)]. 
Such concentration of Berry curvature leading to large BCD was also earlier explored in TaAs-family of Weyl semi-metals~\cite{zhang2018berry}. 
On a similar note, Janus monolayers of Mo-based transition metal carbides (or MXenes) \ch{Mo2COX} (X=S, Se, Te; Cl, Br, I) also exhibit BCD comparable to Janus TMDCs, with BCD values close to 3 \AA{} in the case of \ch{Mo2COTe}~\cite{karmakar2023giant}.

A second approach is the application of an out-of-plane external electric field, as proposed by You \textit{et al.}, where the application of an external field greatly enhances the BCD associated with 1T' \ch{MoTe2} and \ch{WTe2}~\cite{you2018berry,zhang2018electrically}. 
These two-dimensional topological insulators are also known to undergo an electric-field-driven phase transition to normal insulators. The field-induced closing and opening of band gap leads to a large magnitude of BCD and NLH current at the phase transition point, with the sign of BCD inverted across phase transition~\cite{zhang2018electrically}. 
Such a displacement field breaks inversion and allows the systematic control and tuning of the NLH current, and also helps identify topological phase transitions in materials.
In comparison, the T$_d$ phase, without the two-fold screw rotational symmetry, displays a larger BCD than the pristine T' phase, as calculated for 1T$_d$ \ch{WTe2}~\cite{you2018berry,zhang2018electrically}, although the application of electric field largely enhances the BCD present in 1T' phase [Fig.~\ref{fig:theory_wte2} (a)-(b)]. 

Another proposal is the construction of bilayers, as in the case of bilayer 2M \ch{WS2}~\cite{joseph2021topological}. 
This recently discovered monoclinic phase~\cite{fang2019discovery} has inversion symmetry in the bulk form but is non-centrosymmetric as a bilayer due to the stacking arrangement of 1T' monolayers of \ch{WS2}, with only a single mirror plane. The nonlinear Nernst current, a second-order response to a temperature gradient without a magnetic field, is attributed to Berry curvature and its gradient near the Fermi surface. Bilayer WTe$_2$, an inversion-broken type-II Weyl semimetal with chiral Weyl fermions, exhibits a nonlinear anomalous Nernst effect without any magnetic field, which can be efficiently switchable by controlling spin-orbit coupling effects~\cite{zeng2019nonlinear,xu2000vortex}.

Similarly, few-layer \ch{WTe2} has also been explored for NLH effect~\cite{du2018band,wang2019ferroelectric}. 
It was also found that the BCD changes sign across the ferroelectric transition in trilayer \ch{WTe2}~\cite{wang2019ferroelectric}. 
Twisted bilayer \ch{WSe2} exhibits large NLH response under the application of strain which breaks the three-fold rotation symmetry in the system~\cite{hu2022nonlinear}. 
The construction of twisted bilayer \ch{WTe2} ($\theta=29.4\degree$) was further proposed to have a giant NLH current with BCD of the order of $\sim1500$\AA{} due to the presence of a large number of band crossings and band inversions around Fermi level~\cite{he2021giant} [Fig.~\ref{fig:theory_wte2} (c)-(d)]. 
Besides TMDC bilayers, two-dimensional van der Waals (vdW) heterostructures are also possible platforms to measure NLH effect since they naturally break spatial inversion symmetry. Strain tunable BCD was observed in 1T' \ch{MoSe2}/\ch{WSe2} bilayer with values comparable to homogeneous multilayers~\cite{jin2021strain} [Fig.~\ref{fig:theory_wte2} (e)-(f)]. 
Buckled Bi(110) monolayer was predicted to exhibit ferroelectricity-induced BCD upon both doping and the presence of \ch{NbSe2} substrate, coexisting with persistent spin texture~\cite{jin2021enhanced}.
Construction of MXene heterostructures also has a similar result, as seen in T-\ch{Mo2C}/H-\ch{Mo2C}, which notably also displays Ising superconductivity~\cite{zhao2023nonlinear}. 
A recent addition to the class of two-dimensional materials predicted to display NLH effect is the family of composition-tunable materials \ch{Nb_{2n+1}Si_{n}Te_{4n+2}}~\cite{zhao2023berry}. The BCD is maximum for $n=1$, i.e., \ch{Nb3SiTe6} and decreases with increasing $n$, vanishing at $n=\infty$ due to the presence of an extra glide mirror symmetry in the system. 


Apart from two-dimensional materials, three-dimensional Weyl semimetals \cite{bharti2023massless} have also been predicted to exhibit large NLH currents due to the presence of tilted Dirac cones~\cite{zhang2018berry, zeng2021nonlinear}. 
In this respect, type-I (TaAs-type; (Ta, Nb)(As, P)) and type-II (\ch{MoTe2}-type; (Mo, W)\ch{Te2}) Weyl semimetals (WSM) were investigated for BCD. 
Weyl points induce large BCD due to the possible tilting of bands and concentration of Berry curvature.
Therefore, type-II WSMs have been proposed to exhibit larger NLH effect than type-I. 
TaAs-type WSMs belong to $C_{4v}$ point group with two mirror planes $M_x$ and $M_y$. The symmetry restrictions are more relaxed in three dimensions compared to two dimensions, as we discussed previously. 
This leads to two non-zero components in the BCD tensor with the relation $D_{xy} = -D_{yx}$~\cite{zhang2018berry}.
\ch{MoTe2}-type WSM has $C_{2v}$ point group leading to two independent tensor components of BCD -- $D_{xy}$ and $D_{yx}$~\cite{zhang2018berry,singh2020engineering} [Fig.~\ref{fig:theory_bulk} (a)-(b)].
It was also proposed that the construction of TaAs slabs significantly enhances the BCD of the material at the Fermi level~\cite{pang2023tuning}.

Strain-engineered BCD was recently demonstrated in type-II Weyl HgTe~\cite{chen2019strain}, where an in-plane strain can lead to distinct topological phases -- compressive strain leads to a type-I WSM (with a small tilt of bands), and tensile strain results in a topological insulator. 
BCD and NLH effect were also proposed to be used to probe the pressure-driven topological phase transitions -- layered polar material BiTeI goes from a trivial insulator to a topological insulator through a Weyl semimetallic phase under pressure~\cite{facio2018strongly}. A sharp peak in BCD was found at the phase boundary between insulating and Weyl semimetallic phase, followed by a change in the sign of BCD [Fig.~\ref{fig:theory_bulk} (d)]. 
Detection of polarization direction in polar (semi)metals with NLH effect was proposed by investigating the structural phase transition between polar T$_d$ and non-polar 1T' phase of bulk \ch{MoTe2}~\cite{singh2020engineering}. 
NLH signal is predicted to vanish in the 1T' phase while it remains non-zero in the T$_d$ phase. 

Beyond the TMDC family, BCD induces circular photogalvanic effect (CPGE) and NLH effect in p-doped trigonal Te~\cite{tsirkin2018gyrotropic}.  
The two-dimensional surface band structure of three-dimensional topological insulators of the \ch{Bi2X3} (X= Se, Te) family is known to have an odd number of Dirac cones~\cite{chen2009experimental,xia2009observation}. The presence of hexagonal warping (similar to the strain-induced warping of two-dimensional systems, discussed earlier) in the surface band structure leads to finite BCD in \ch{Bi2Te3}, and hence the system exhibits NLH response~\cite{yar2022nonlinear}. 
The pronounced NLH response in the system was explained as a result of the interplay between disorder scattering and hexagonal warping, but without warping the NLH effect will not exist. Moreover, Hopf insulators~\cite{deng2013hopf} in three dimensions represent a distinct class of topological phases going beyond the conventional tenfold-way classification \cite{schnyder2008classification}. During the critical transition between two rotational symmetry-based Hopf insulators, band-touching points form point Berry dipoles with overlapping opposite Berry dipole charges. This overlapping provides a unique Berry curvature distribution and specific quantization of Berry flux near these transitions, which relates the nonlinear effect with Hopf invariants~\cite{zhuang2023extrinsic}.
A recent study on antiferromagnetic CuMnAs reveals the presence of a time-reversal-odd second-order conductivity, which was termed an intrinsic NLH effect (INLHE)~\cite{wang2021intrinsic}. This INLHE is different from NLH arising due to BCD (also called the intrinsic contribution to NLH), which is a time-reversal-even quantity, forbidden in the $PT$-symmetric antiferromagnetic systems. The INLHE is independent of the relaxation time $\tau$, while BCD-induced NLH is proportional to $\tau$.
There have been several other materials proposed to exhibit BCD and NLH effect based on (lack of) symmetry alone -- the topological phase of inversion breaking \ch{ZnGeSb2}~\cite{sadhukhan2022pressure}, rotational symmetry broken supertwisted \ch{WS2}~\cite{ci2022breaking}, two-dimensional Dirac semimetals of SbSSn family~\cite{jin2020two}, and Gd-intercalated graphene multilayers~\cite{kolmer2022highly} are a few such materials. 
Different candidate materials predicted to display BCD and the NLH effect, along with the different strategies used to obtain large and tunable BCD has been listed in Table~\ref{tab:materials_candidate}.
\begin{longtable}{c| c |c |c}
     \caption{A summary of the candidate materials, their dimensionality, associated symmetry, and the strategy to obtain nonzero BCD. } 
     \label{tab:materials_candidate} \\
     \hline
     \hline
       Candidate material & Dimension & Symmetry & Notes\\ 
      \hline
      \hline
       \shortstack{Graphene \\
(Monolayer~\cite{battilomo2019berry} , \\ bilayer~\cite{yoon2017broken}, \\ twisted \\ bilayer~\cite{pantaleon2021tunable, zhang2022giant, arora2021strain, pantaleon2022interaction})} & 2D & \shortstack{Inversion; \\ high symmetry; \\ broken using substrate/\\ electric field/ strain} & -- \\ [0.6ex]
    \hline
        \shortstack{Buckled honeycomb \\(Silicene, Germanene, \\ Stanene) ~\cite{bandyopadhyay2022electrically}} & 2D & \shortstack{Inversion; C$_3$; \\ broken using \\ electric field and \\ uniaxial strain} & \shortstack{Large value of BCD \\ at critical field\\ $\sim$ nm}\\ [0.5ex]
       \hline
        Phosphorene~\cite{bandyopadhyay2023berry} & 2D & \shortstack{Inversion; \\ broken using staggered \\ onsite potential/ \\ SnSe substrate} & Strain-tuneable BCD\\ [0.5ex]
        \hline
        \shortstack{1H \ch{MX2} \\
(M - Mo, W, Nb, Ta; \\ X - S, Se) ~\cite{son2019strain, you2018berry, xiao2020two}} & 2D & \shortstack{No inversion; \\ Strain-induced BCD} & \shortstack{Piezoelectric-like \\ behaviour in metallic \\ systems}\\[0.5ex]
        \hline
         1H MoSSe~\cite{zhou2020highly} & 2D & \shortstack{No inversion; \\ strain-induced BCD} & -- \\ [0.5ex]
        \hline
        \shortstack{1T’ \\ WSeTe, WSTe~\cite{joseph2021tunable}} & 2D & \shortstack{No inversion; \\ low symmetry} & -- \\ [0.5ex]
        \hline
         \shortstack{1T’ \\ \ch{MoTe2}, \ch{WTe2}~\cite{you2018berry,zhang2018electrically}} & 2D &\shortstack{Inversion; \\ broken with out-of-plane \\ electric field}  & -- \\ [0.5ex]
         \hline
          $T_d$ \ch{WTe2}~\cite{you2018berry,zhang2018electrically} & 2D & \shortstack{No inversion; \\ low symmetries} & -- \\ [0.5ex]
         \hline
          \shortstack{\ch{Mo2COX} \\(X=S, Se, Te; Cl, Br, I)}~\cite{karmakar2023giant} & 2D & No inversion & BCD $\sim 3$ \AA{}\\ [0.5ex]
         \hline
          \shortstack{Bilayer \\ 2M \ch{WS2}~\cite{joseph2021topological}} & 2D & No inversion& -- \\ [0.5ex]
         \hline
          \shortstack{Few-layer \ch{WTe2}}~\cite{du2018band,wang2019ferroelectric} & 2D & No inversion & \shortstack{BCD changes sign\\across ferroelectric\\transition} \\ [0.5ex]
         \hline
          \shortstack{Twisted bilayer \\ \ch{WSe2}~\cite{hu2022nonlinear}} & 2D & \shortstack{No inversion; \\C$_3$ broken by strain} & -- \\ [0.5ex]
         \hline
          \shortstack{Twisted bilayer\\ \ch{WTe2}~\cite{he2021giant}} & 2D & No inversion & \shortstack{Large no. of \\band crossings;\\ BCD $\sim 1500$ \AA{}} \\ [0.5ex]
         \hline
          \shortstack{1T’ \\ \ch{MoSe2/WSe2} bilayer}~\cite{jin2021strain} & 2D & No inversion & Strain-tunable BCD\\ [0.5ex]
         \hline
          \ch{T~ Mo2C/H~ Mo2C}~\cite{zhao2023nonlinear}  & 2D & No inversion  & \shortstack{Co-existing BCD \\ and Ising superconductivity}\\ [0.5ex]
         \hline
          Bi(110) monolayer~\cite{jin2021enhanced} & 2D & \shortstack{No inversion \\ (puckered monolayer)} & \shortstack{Ferroelectricity-induced\\BCD on doping/substrate}\\ [0.5ex]
         \hline
          \ch{Nb_{2n+1}Si_{n}Te_{4n+2}}~\cite{zhao2023berry} & 2D & \shortstack{No inversion; \\ glide mirror at $n=\infty$} & \shortstack{Max BCD \\at $n=1$} \\ [0.5ex]
         \hline
          \shortstack{Type-I~\cite{zhang2018berry, zeng2021nonlinear}/\\Type-II~\cite{ zeng2021nonlinear,singh2020engineering} \\ Weyl semimetals} & 3D & No inversion & \shortstack{Higher BCD in \\ type-II Weyl due to \\ larger tilt}\\ [0.5ex]
         \hline
          \ch{LiOsO3}~\cite{xiao2020electrical} & 3D& No inversion & \shortstack{BCD strongly depends \\on polar distortion}\\
         \hline
         SnTe ~\cite{kim2019prediction,wang2019ferroicity} & 3D & No inversion & \shortstack{Ferrorelectrically-\\ controlled BCD}\\
         \hline
          HgTe~\cite{chen2019strain}  & 3D& No inversion & \shortstack{Strain-tunable \\BCD} \\
         \hline
          BiTeI~\cite{facio2018strongly}  & 3D& No inversion & \shortstack{BCD peaks at topological \\phase boundary}\\
         \hline
         Trigonal Te~\cite{tsirkin2018gyrotropic}  & 3D& No inversion & -- \\ [1.5ex]
         \hline
         \ch{Bi2Te3}~\cite{yar2022nonlinear}  & 3D& & \shortstack{Disorder scattering \\ 
         and hexagonal warping}\\
         \hline
         Hopf insulators~\cite{zhuang2023extrinsic} & 3D & & \shortstack{NLH effect \\ with Hopf invariants}\\
         \hline
          CuMnAs~\cite{wang2021intrinsic} & 3D & $PT$ symmetry & Intrinsic NLH effect \\ [1.5ex]
         \hline
\end{longtable} 


\section{Experimental advances}

\begin{figure*}[t]
    \centering
    \includegraphics[width=14cm]{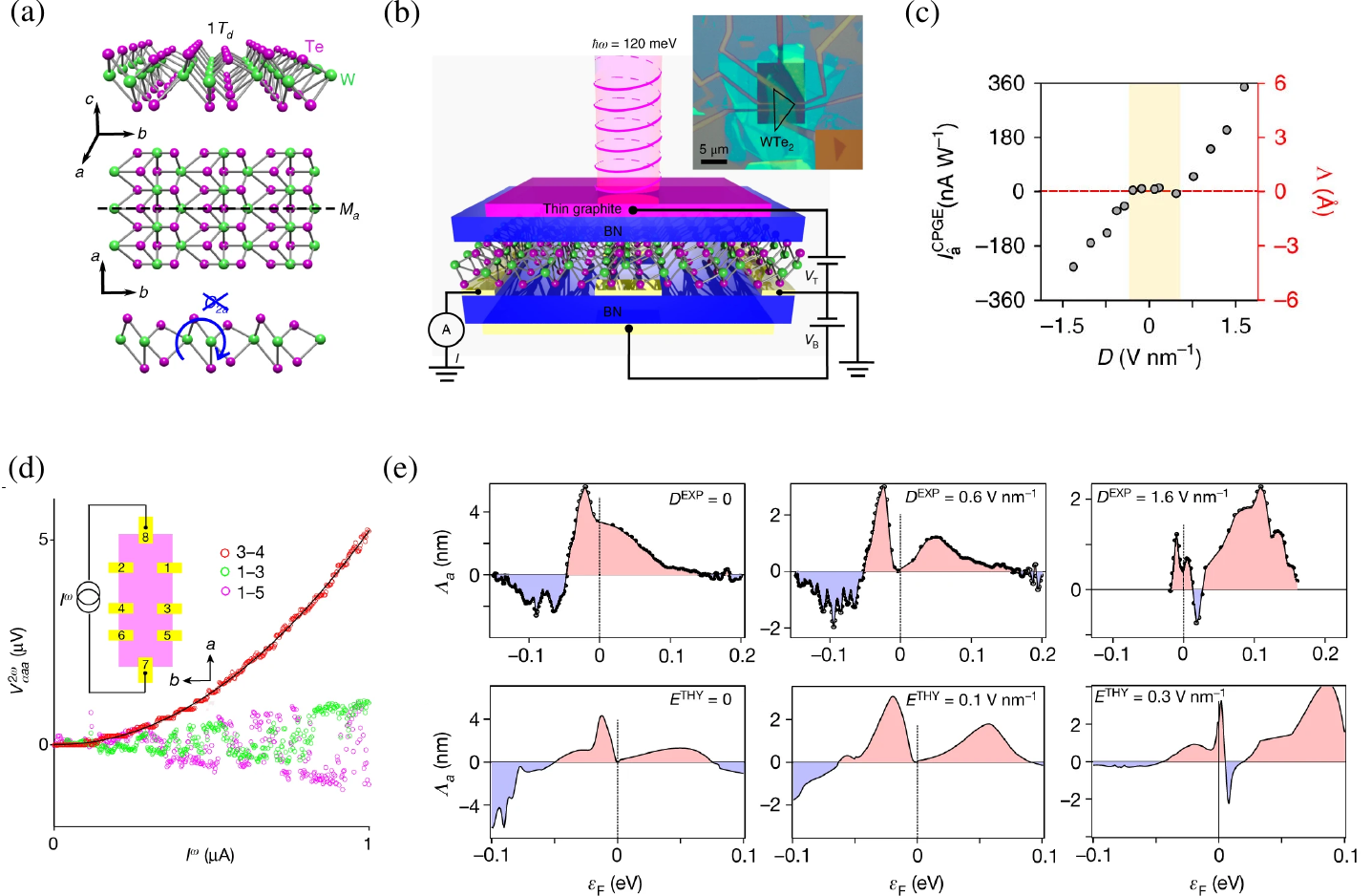}
    \caption{\textbf{Experimental measurement of Berry curvature dipole and NLH effect in \ch{WTe2}}. (a) 1T$_d$ phase of monolayer \ch{WTe2} has broken inversion with a single mirror plane, ideal for the detection of NLH current. (b) Schematic experimental setup to detect mid-infrared circular photogalvanic effect (CPGE). This CPGE is induced due to the presence of BCD in the material. The inset shows the optical image of the device. (c) The measured CPGE current as a function of displacement field (left axis) and estimated BCD (right axis). Reproduced with permission from~\cite{xu2018electrically}. (d) The measured NLH current in bilayer \ch{WTe2}. Inset marks the different electrodes between which the current was measured. The transverse voltage is nonlinear with respect to the input current and dominates the response. (e) Experimental (top) vs theoretical estimation (bottom) of BCD ($\Lambda_a$) at different displacement fields for bilayer \ch{WTe2}. Reproduced with permission from~\cite{ma2019observation}.}
    \label{fig:expt_wte2}
\end{figure*}

This section details the major experimental observations of BCD and NLH effects in different materials. 
As we discussed previously, the maximum allowed symmetry for non-zero BCD in two-dimensional materials is a mirror plane. This symmetry constraint leads to a restricted number of materials that can possibly display these NLH properties. Among the candidate materials originally proposed by Sodemann and Fu~\cite{sodemann2015quantum} are two-dimensional TMDCs, and the first experimental detection of BCD and the resulting NLH effect was also performed in this class of materials. 

Among the first experiments was the indirect evidence of the presence of BCD in monolayer \ch{WTe2}~\cite{xu2018electrically}, a known quantum spin Hall (QSH) insulator~\cite{qian2014quantum}. They considered the 1T$_d$ monolayer of \ch{WTe2}, isolated from the inversion-breaking bulk T$_d$ lattice. 
This phase is similar to the 1T' phase of TMDCs -- a distorted octahedral metal-chalcogen coordination -- but differs from it because it lacks a two-fold screw rotation symmetry $C_{2a}$ [Fig.~\ref{fig:expt_wte2} (a)]. The combination of $C_{2a}$ and the mirror symmetry gives rise to an inversion symmetry in the 1T' phase. The weakly broken screw rotation renders the 1T$_d$ phase non-centrosymmetric, making it ideal for detecting BCD. 
A CPGE is induced with a normally incident light in \ch{WTe2} by the interband transition across the inverted QSH gap. This CPGE was observed to be enhanced and tunable under an external, out-of-plane displacement field and was attributed to the presence of a non-zero BCD in the system. 

The first direct measurement of NLH effect was also carried out on two-dimensional \ch{WTe2}, where a bilayer was considered to satisfy the required crystalline symmetry conditions for a non-zero BCD~\cite{ma2019observation}, since the monolayer is "approximately" inversion symmetric. A clear quadratic response was observed with double the input current frequency along the transverse direction, which also dominated the longitudinal response [Fig.~\ref{fig:expt_wte2} (d)]. The combined quadratic current-voltage response and the 2$\omega$ frequency confirm the second-order nature of the output voltage. The NLH angle was also observed to be close to 90$^{\circ}$. 
The BCD determined from the experimentally observed second-order response was found to be in good agreement with the results from first-principles calculations, thereby validating that the NLH effect in \ch{WTe2} is induced by the BCD present [Fig.~\ref{fig:expt_wte2} (e)]. The non-uniform distribution of Berry curvature leading to the formation of BCD was attributed to the presence of layer-polarized Dirac fermions in bilayer \ch{WTe2}.

A similar NLH effect was also measured in few-layer T$_d$ \ch{WTe2} through angle-resolved electrical measurements~\cite{kang2019nonlinear}, where the corresponding BCD obtained was comparable to previously predicted values.
The study also concluded, through temperature-dependent measurements, that the observed NLH effect in few-layer \ch{WTe2} has contributions from both intrinsic BCD and extrinsic spin-dependent scattering.
Similarly, in T$_d$ \ch{MoTe2}, while NLH effect had major contributions from BCD in thin layers, it was dominated by extrinsic charge scattering for thicker layers~\cite{ma2022growth}. 

Apart from transition metal ditellurides, which exhibit naturally non-centrosymmetric low-symmetry structures, other members of the TMDC family have also been explored for the presence of NLH effect and BCD. Applying uniaxial strain to break the $C_{3v}$ rotation axis leads to the formation of a strain tunable BCD in 1H \ch{WSe2}~\cite{qin2021strain}, comparable to that previously measured for \ch{WTe2}~\cite{ma2019observation}.
Similar strain engineering was previously performed for monolayer \ch{MoS2}~\cite{son2019strain} where the strain-induced BCD leads to the formation of valley orbital magnetization as a response to an in-plane electric field. 
Another engineered structure to break symmetry is the construction of twisted bilayers, as done in the case of twisted bilayer \ch{WSe2}~\cite{huang2023giant}, where a giant peak in the NLH current was observed at the half-filling of moir\'e bands. 
Construction of bilayer T$_d$-\ch{WTe2}/monolayer H-\ch{WSe2} moir\'e heterostructure (superlattice) leads to a switchable NLH transport by ferroelectricity~\cite{kang2023switchable}.

Graphene is another two-dimensional material widely explored for the presence of NLH effect. The high symmetry of the lattice has been broken by the construction of corrugated bilayers~\cite{ho2021hall}, twisted bilayers~\cite{huang2023intrinsic,duan2022giant} and double-bilayers~\cite{sinha2022berry,zhong2023effective}, moir\'e superlattices~\cite{he2022graphene} of graphene. The origin of NLH effect in graphene has been actively debated to be due to the presence of BCD or the disorder-induced skew-scattering. 
The transverse nature of NLH effect present in artificially corrugated graphene is suggestive of induced BCD~\cite{ho2021hall}. On the other hand, observations in graphene moir\'e superlattice show that the NLH effect is due to the skew-scattering of the chiral Bloch waves, with the nonlinear conductivities five orders of magnitude larger than those in \ch{WTe2}~\cite{he2022graphene}.
While one study on twisted double-bilayer graphene (TDBG) concludes that the nonlinear current is due to the presence of large BCD due to strain~\cite{sinha2022berry}, a more recent studys observed that BCD dominates the NLH effect in TDBG, but near the band edge, a controllable transition to skew-scattering-dominated response is possible. 
Both disorder-induced skew-scattering~\cite{duan2022giant} and BCD~\cite{huang2023intrinsic} have been put forth as the reason behind nonlinear conductivity in twisted bilayer graphene. 
The first observation of NLH effect in an organic conductor was in the weak charge-ordered state of $\alpha$-\ch{(BEDT-TTF)2I3}~\cite{kiswandhi2021observation}, a multilayer two-dimensional Dirac fermion system with a pair of Dirac cones obtained at high pressure. 

Besides two-dimensional materials, three-dimensional Weyl and Dirac semimetals and topological crystalline insulators (TCI) have been proposed as possible candidates to detect NLH effect. This was also experimentally confirmed, with NLH voltage measured in single crystals of Dirac materials \ch{Cd3As2}~\cite{shvetsov2019nonlinear} and \ch{BaMnSb2}~\cite{min2023strong}, Weyl semimetals \ch{WTe2}~\cite{shvetsov2019nonlinear} and \ch{TaIrTe4}~\cite{kumar2021room}, and TCI Pb$_{1-x}$Sn$_x$Te~\cite{zhang2022pst,nishijima2023ferroic}.
Among these, room temperature NLH current was measured for the spin-valley locked Dirac state of \ch{BaMnSb2}~\cite{min2023strong} and Weyl semimetallic states of \ch{TaIrTe4}~\cite{kumar2021room}, and more recently in elemental Bismuth thin films~\cite{makushko2023tunable}, focused ion beam deposited Pt~\cite{min2023colossal}, and topological semimetal GeTe~\cite{orlova2023gate}.
In bulk samples of T$_d$ \ch{MoTe2} and \ch{WTe2}, a large out-of-plane NLH effect was measured, which was predominantly the result of asymmetric electron scattering~\cite{tiwari2021giant}.
Such nonlinear currents as a result of skew-scattering (in the absence of BCD) have also been observed in the topological insulator \ch{Bi2Se3}~\cite{he2021quantum} and type-II Dirac semimetal 1T \ch{CoTe2}~\cite{hu2023terahertz}.
Later, Yar \textit{et. al.}~\cite{yar2022nonlinear} proposed that hexagonal warping also plays a crucial role in the class of \ch{Bi2X3} materials, whose interplay with disorder scattering lead to pronounced NLH response in \ch{Bi2Te3} (see ~\ref{sec:materials}).
A more recent study on bulk T$_d$ \ch{WTe2} has an effective inversion symmetry in the transport plane due to the combination of a mirror plane and a glide mirror plane, broken with the application of an electric field. This field-induced BCD provides a promising method to generate and control BCD in material systems~\cite{ye2023control}. 

The giant SHG observed in two-dimensional Te was also attributed to the presence of BCD in the system~\cite{fu2023berry}. 
Second-order non-linear transport in the superconducting state was measured in trigonal \ch{PbTaSe2}, with the non-linear signal significantly enhanced around the superconducting transition~\cite{itahashi2022giant}.
The presence of intrinsic NLH effect in antiferromagnetic materials, induced by the quantum metric dipole (BCD vanishes for such systems, see \ref{sec:materials}), was, very recently, experimentally discovered in \ch{MnBi2Te4}-black phosphorous heterostructure~\cite{gao2023quantum}. 
Recent experiments also measured a \textit{third-order} NLH effect in T$_d$ \ch{MoTe2}~\cite{lai2021third}, \ch{TaIrTe4}~\cite{wang2022room}, and \ch{Cd3As2}~\cite{zhao2023gate}. Here the non-linear response is connected to the Berry-connection polarizability tensor, as we discussed.
Table~\ref{tab:expt_materials} summarizes the different materials and the major factors contributing to the measured NLH effect.
\begin{longtable} {c|c|c|c}  
  \caption {A summary of the experimentally investigated materials, their dimensionality, NLH contribution, and key notes. } \label{tab:expt_materials} \\
  \hline
  \hline
      Material  &  Dimension & NLH contribution     & Notes   \\
   \hline
   \hline
      \shortstack{ Monolayer \\ T$_d$ \ch{WTe2}~\cite{xu2018electrically}}   & 2D  & Intrinsic (BCD) &  -- \\
   \hline
       \shortstack{ Bilayer \\ 1T’ \ch{WTe2}~\cite{ma2019observation} }  & 2D  & Intrinsic (BCD)     &  -- \\
   \hline  
       \shortstack{Few-layer \\ T$_d$ \ch{WTe2}~\cite{kang2019nonlinear}}  & 2D  & \shortstack{Intrinsic (BCD) and extrinsic \\spin-dependant scattering} & -- \\
    \hline
      T$_d$ \ch{MoTe2} ~\cite{ma2022growth}    & 2D  & \shortstack{Dominant charge \\scattering in thicker \\layers}   &  --   \\
    \hline
     1H \ch{MoS2} ~\cite{son2019strain},\ch{WSe2} ~\cite{qin2021strain}    & 2D  & Intrinsic  & \shortstack{C$_3$ broken by \\ uniaxial strain}  \\
    \hline
      Twisted bilayer \ch{WSe2} ~\cite{huang2023giant}   & 2D  & Intrinsic  & \shortstack{Giant NLH peak\\ at half-filling of\\ moir\'e bands}  \\ [0.5ex]
    \hline
      \shortstack{T$_d$-\ch{WTe2}/ H-\ch{WSe2} \\ superlattice ~\cite{kang2023switchable}}   & 2D  & Intrinsic & \shortstack{Ferroelectrically \\switchable NLH effect} \\
   \hline
     Corrugated Graphene ~\cite{ho2021hall}    & 2D  & Intrinsic    &   --    \\ [0.5ex]
   \hline
     \shortstack{ Graphene moir\'e \\ superlattice ~\cite{he2022graphene}}   & 2D  & \shortstack{skew-scattering of \\the chiral Bloch waves}  &    --  \\
   \hline
     \shortstack{Twisted \\double-bilayer \\graphene ~\cite{sinha2022berry}$^a$ \\~\cite{zhong2023effective}$^b$} & 2D  & \shortstack{$^a$Intrinsic~\\ $^b$Skew-scattering \\at band edge}    & $^a$Large BCD due to strain       \\
   \hline
      \shortstack{Twisted bilayer\\ graphene ~\cite{huang2023intrinsic}$^a$ \\~\cite{duan2022giant}$^b$}    & 2D  & \shortstack{$^a$Intrinsic~\\$^b$Disorder induced \\ skew-scattering}   &  --    \\
   \hline
      $\alpha$-\ch{(BEDT-TTF)2I3}~\cite{kiswandhi2021observation}     & 2D  &  Intrinsic  & \shortstack{NLH effect in an organic \\ conductor} \\
   \hline
     \shortstack{Dirac Materials \\ \ch{Cd3As2}~\cite{shvetsov2019nonlinear} \\ \ch{BaMnSb2}~\cite{min2023strong}*}  & 3D  & *Intrinsic & \multirow{2}{*}{\shortstack{*Room temperature \\ NLH effect}}  \\
    \cline{1-3}
    \shortstack{ Weyl semimetals \\ \ch{WTe2}~\cite{shvetsov2019nonlinear} \\ \ch{TaIrTe4}~\cite{kumar2021room}*} & 3D  & *Intrinsic &    \\
   \hline
     Pb$_{1-x}$Sn$_x$Te~\cite{zhang2022pst,nishijima2023ferroic} & 3D   & Intrinsic &  --   \\ [0.5ex]
   \hline
   Bi thin films ~\cite{makushko2023tunable}   & 3D   & \shortstack{Disorder-induced \\ scattering}  &  \multirow{3}{*}{\shortstack{Room temperature \\ NLH effect}}  \\ [0.5ex]
    \cline{1-3}
   Pt~\cite{min2023colossal}    & 3D   & \shortstack{Disorder-induced \\ scattering}   &  --   \\ [0.5ex]
    \cline{1-3}
   GeTe~\cite{orlova2023gate}     & 3D   &  Intrinsic  &     \\ [0.5ex]
    \hline
    \shortstack{T$_d$ \ch{MoTe2},\\ \ch{WTe2}~\cite{tiwari2021giant}}  & 3D   & \shortstack{Asymmetric \\ electron scattering}   &  --  \\
     \hline
   \ch{Bi2X3} (X=Se,Te)~\cite{he2021quantum,yar2022nonlinear}     & 3D   & \shortstack{Hexagonal warping \\ and disorder scattering}      &   --   \\ [0.5ex]
    \hline
    1T \ch{CoTe2}~\cite{hu2023terahertz}     & 3D   &  skew-scattering   &  --     \\ [0.5ex]
    \hline
   \ch{PbTaSe2} ~\cite{itahashi2022giant}   &  3D   &  Scattering      & \shortstack{Enhanced NLH \\signal around \\ superconducting \\transition} \\
    \hline
    \shortstack{\ch{MnBi2Te4}/ \\ black phosphorous ~\cite{gao2023quantum}}     & 3D   & INLHE   & Antiferromagnetic  \\
     \hline
   T$_d$ \ch{MoTe2}~\cite{lai2021third}       & 3D   & \multirow{3}{*}{\shortstack{Berry connection \\ polarizability}}     &\multirow{3}{*}{\shortstack{Third-order \\ NLH effect}}   \\ [0.5ex]
  \cline{1-2}
     \ch{TaIrTe4}~\cite{wang2022room}       & 3D   & &      \\ [0.5ex]
 \cline{1-2}
  \ch{Cd3As2}~\cite{zhao2023gate} &3D   & &  \\ [0.5ex]
   \hline
\end{longtable}


\section{Future directions and outlook}


In the preceding sections, we explored the different facets of the NLH effect. Next, we highlight the very recent developments in closely related directions, which are pushing the boundaries in this rapidly expanding field. 
Quantum geometric properties of electron wave functions lead to various electronic transport phenomena~\cite{papaj2019magnus,nagaosa2010anomalous,gao2021layer,xiao2010berry}. In the field of optics, the Berry phase is crucial in determining various fascinating responses, including photogalvanic effects, circular photogalvanic mechanisms, and some other nonlinear optical phenomena~\cite{sodemann2015quantum,de2017quantized,watanabe2021chiral}. We already discussed that key factors such as the BCD and quantum metric play a predominant role in determining nonlinear optical responses \cite{bharti2023role}. Recent interest in the band geometric properties of quantum materials has introduced new aspects of nonlinear phenomena, such as nonreciprocal currents, Hall effects, and probing crystallographic details from transport~\cite{tokura2018nonreciprocal,carvalho2019nonlinear,yin2014edge}. However, the interplay between band geometric quantities contributing to photogalvanic effects, second harmonic response, and higher harmonic generation~\cite{rostami2018nonlinear,you2018berry,golub2018circular,ma2019nonlinear,okamura2020giant,yoshikawa2019interband,tan2019upper,wang2020electrically,fei2020giant,zou2020alternating,kaplan2022twisted,gao2021current,wei2021electric,chaudhary2022shift,shi2021geometric,hamara2023ultrafast} are yet to be fully understood. To address this challenge, Bhalla \textit{et al.} have extended the theory of second harmonic generation to encompass Fermi surface effects in metallic systems and consider a finite scattering timescale~\cite{bhalla2022resonant}. In doped materials, the presence of the Fermi surface and Fermi sea introduces resonances in all second harmonic terms. Notably, two distinct contributions to the second harmonic signal are proposed -- a double resonance arising from the Fermi surface and a higher-order pole due to the Fermi sea. The quantum geometric classification mentioned above has been critically explained in the monolayer of the time-reversal symmetric Weyl semimetal WTe$_2$ and the parity-time reversal symmetric topological antiferromagnet CuMnAs.

In a recent work, Ahn \textit{et al.} have investigated the low-frequency properties of the bulk photovoltaic effect in topological semimetals~\cite{ahn2020low}, exploring its potential for terahertz photodetection~\cite{kraut1979anomalous,von1981theory,morimoto2016topological,de2017quantized,liu2020semimetals}. The analysis of the second-order optical conductivity in point-node semimetals~\cite{ahn2023topological} with tilted cones reveals divergent behaviors at Dirac and Weyl points. In particular, under circularly polarized light, conductivity scales as $\omega^{-2}$ and $\omega^{-1}$ in two and three dimensions, respectively. The bulk photovoltaic effect essentially originates from two mechanisms -- the transition of electron position and the transition of electron velocity during optical excitation. This results in two types of photocurrents, known as the shift current and the injection current~\cite{sipe2000second}. An analysis of two-band models has demonstrated that the injection current is primarily influenced by the quantum metric and Berry curvature, while the shift current is governed by the Christoffel symbols near gap-closing points in semimetals. These results regarding the shift and injection photocurrent conductivities have been experimentally verified in the case of antiferromagnetic MnGeO$_3$~\cite{xu2020high} and ferromagnetic PrGeAl~\cite{chang2018magnetic} systems, as prototypical examples of magnetic Dirac and Weyl semimetals. As we have seen, understanding how electronic systems respond to static electromagnetic fields, in the case of quantum and anomalous Hall effects, relies on the well-established geometry of quantum states. However, linking this quantum geometry to resonant optical responses has been challenging due to the involvement of pairs of states in optical transitions. Existing geometric properties are defined for a single state, limiting understanding to two-level systems. For that purpose, a general theory of Riemannian geometry for resonant optical processes, applicable to arbitrarily high-order responses, has been proposed~\cite{ahn2022riemannian,bouhon2023quantum,jankowski2023optical}. The study suggests that optical responses can be seen as manifestations of the Riemannian geometry of quantum states. In particular, the third-order photovoltaic Hall effects of Dirac and Weyl fermions are closely related to the Riemann curvature tensor. Furthermore, this theoretical framework extends beyond electromagnetic responses, offering insights into other geometric responses such as the photovoltaic thermal Hall and Seebeck effects. It is also feasible to explore real-space quantum geometry by designating momentum operators as tangent vectors. These exciting avenues remain an open problem for future research in this field.

Recent developments have been made in understanding nonlinear planar Hall effects. Kheirabadi \textit{et al.} have explored the quantum nonlinear planar Hall effect in bilayer graphene subjected to a constant in-plane magnetic field~\cite{kheirabadi2022quantum}. The introduction of the magnetic field, breaking time-reversal symmetry, results in a charge current as a second-order response to an external electric field, influenced by the BCD in momentum space. The nonlinear planar Hall effect, arising from the anomalous velocity, stems from the orbital impact of the in-plane magnetic field on electrons in bilayer graphene, independently of spin-orbit coupling. We note that, some of the nonmagnetic polar and chiral crystals such as Janus monolayer MoSSe~\cite{zhang2017janus,lu2017janus} exhibit intrinsic nonlinear planar Hall effect~\cite{huang2023intrinsicother}. Furthermore, a different approach has been adopted by Hayami \textit{et al.} to generate spin currents in collinear antiferromagnets (AFMs) with parity-time (PT) symmetry~\cite{hayami2022nonlinear}. The spin current has been achieved in terms of the nonlinear spin Hall effect, where spatial inversion (P) and time-reversal (T) operations play crucial roles. 
Furthermore, electric field induced nonlinear spin currents can be observed in centrosymmetric magnets such as single-layer MnBi$_2$Te$_4$~\cite{xiao2022intrinsic}. It is worth mentioning that other centrosymmetric magnets, for instance, 1T-MnSe$_2$~\cite{o2018room}, CrI$_3$~\cite{huang2017layer}, and 1T-VSe$_2$~\cite{bonilla2018strong} will also host a similar nonlinear electric spin generation. We note that the symmetry base indicators for this nonlinear spin current are different from the conventional linear spin current~\cite{naka2019spin,yuan2020giant}. The microscopic origin of a pure second-order spin current in metals is spin-dependent BCD. This BCD is induced by an effective spin-dependent hopping mechanism within AFM ordering. There can be a significant response, particularly near the AFM phase transition accompanying emergent uniform magnetic toroidal dipole~\cite{spaldin2008toroidal,yatsushiro2022analysis}. By eliminating the need for SOC, uniform magnetization, and spin-split band structures, this BCD-induced mechanism expands the range of potential materials for the development of next-generation spintronic devices centered around AFMs. Further, disorder plays a crucial role in the nonlinear anomalous Hall effect, with the skew scattering and side-jump effects significantly overshadowing the intrinsic impact associated with the BCD~\cite{atencia2023disorder}. Tetrahedral semiconductors, such as GaAs zincblende crystals, provide a distinctive path for investigating nonlinear electrical responses through the breaking of inversion symmetry. This effect is especially pronounced in GaAs, characterized by a significant spin-orbit interaction. The impact is further magnified in spin-3/2 hole systems, setting them apart from spin-1/2 electrons~\cite{shanavas2016theoretical,durnev2014spin, marcellina2018electrical,liu2018strong,wang2021optimal,cullen2021generating,hendrickx2018gate}.

The NLH effect is predicted to lead to potential applications, in addition to fundamental science. Terahertz technology is rapidly advancing with applications in security, quality control, medical imaging, and autonomous vehicle sensing~\cite{tonouchi2007cutting,sizov2010thz}. Terahertz-band communication is pivotal for future wireless networks. However, achieving fast, sensitive, and broad terahertz detection at room temperature poses challenges. Conventional detectors need cryogenic cooling and operate slowly, while alternatives like Schottky diodes with antennas have cutoff frequency limitations below 1 THz. Zhang and Fu have recently predicted a state-of-the-art terahertz detection method based on the intrinsic NLH effect in quantum materials~\cite{zhang2021terahertz}. The approach utilizes the quadratic transverse current-voltage characteristic in these materials to directly detect terahertz radiation, eliminating the need for any junction region. This particular type of "Hall rectifier" offers intrinsic full-wave rectification with a large responsivity at zero bias and a high cutoff frequency spanning from subterahertz to tens of terahertz~\cite{isobe2020high}. As we discussed, beyond the BCD, other factors such as skew scattering and side jump contribute to the NLH effect. This is because, in specific crystal structures, second-order nonlinear transport is influenced by the inherent chirality of the Bloch electron wavefunction. Moreover, in the ballistic transport regime, the NLH conductance is determined by the integral of Berry curvature over half of the Fermi surface~\cite{papaj2019magnus}. Regardless of their origins, these intrinsic second-order nonlinearities in quantum materials can be exploited for long-wavelength photodetection by converting oscillating electric fields into direct current. A recent breakthrough revealed a room temperature NLH effect in the type II Weyl semimetal TaIrTe$_4$, enabling radiofrequency rectification~\cite{kumar2021room}. Additionally, room temperature terahertz detection and imaging were achieved using the second-order nonlinear response of topological surface states in the Dirac semimetal PdTe$_2$~\cite{guo2020anisotropic}. These promising findings are poised to inspire future advancements in terahertz/infrared technology centered on Hall rectifiers. Analyzing the response of a circular drive serves as an effective means of detecting the topology of the lowest-energy Bloch band, correlating with a frequency-dependent probe. This analysis reveals circular dichroism \cite{mandal2023signatures}, driven by differential excitation rates induced by left- and right-circular orientations of a time-periodic drive on a filled band, attributed to the inherent geometric properties of the Bloch bands \cite{sekh2022circular}.

NLH response can efficiently probe the transition between quantum phases, characterized by the breaking of inversion symmetry of the doped tilted Dirac systems~\cite{rostami2020probing}. Moreover, a new nonlinear valley Hall effect has recently been proposed in materials with spatial inversion symmetry~\cite{das2023nonlinear}. To comprehend the origin of this nonlinear valley Hall effect, one must delve into semiclassical electron dynamics, incorporating corrections up to the second order in the electric field~\cite{gao2014field,sundaram1999wave}. This discovery enables control over valley degrees of freedom in centrosymmetric materials, with potential applications in valley-caloritronics. The nonlinear valley Nernst~\cite{dau2019valley} and thermal Hall effects~\cite{chen2022magnon} offer exciting possibilities, extending to bosonic systems. Finally, we note that the family of polar metals, where the coexistence of polarity and metallicity within the same phase is observed, may serve as a promising platform for realizing and understanding these NLH effects~\cite{xiao2020electrical,bhowal2023polar,jager2023universal}. In particular, the BCD may serve as an order parameter of the phase transition in these polar metals. In future, it will be interesting to find other phases for which BCD and its higher order analogs could serve as experimentally-measurable order parameters.

The nonlinear Hall response, in addition to its fundamental interest, also presents numerous practical applications that highlight the remarkable characteristics of emergent quantum materials. One such application lies within the realm of wireless radiofrequency rectification. The room-temperature nonlinear Hall effect manifested in TaIrTe$_4$ has been harnessed to facilitate wireless radiofrequency rectification in a magnetic field and bias-free environment~\cite{kumar2021room}. Moreover, given its second-order characteristic, the second-order nonlinear Hall response presents an avenue for converting oscillating electromagnetic fields into direct current, thus offering a promising method for electromagnetic energy harvesting. Furthermore, the observation of the second-order nonlinear Hall effect in type-II Dirac semimetal CoTe$_2$ under time-reversal symmetry is ascribed to disorder-induced extrinsic contributions on the surface with broken inversion symmetry, enabling room-temperature terahertz rectification devoid of semiconductor junctions or bias voltage~\cite{hu2023terahertz}. We note that for CoTe$_2$, a photoresponsivity exceeding $0.1 \, \mathrm{A \, W^{-1}}$, a response time of approximately $710 \, \mathrm{ns}$, and a mean noise equivalent power of $1 \, \mathrm{pW \, Hz^{-1/2}}$ are attained at room temperature. These results unveil a novel avenue for low-energy photon harvesting through nonlinear Hall-induced nonlinear rectification in strongly spin–orbit-coupled and inversion-symmetry-breaking systems, promising significant advancements in the realm of infrared/terahertz photonics. 

Recently, Zhang \textit{et al.} have explored the method underpinned by the second-order Hall response to rectify the incident terahertz or infrared electric field into a direct current without requiring any diode~\cite{zhang2021terahertz}. In the above case, the photodetector operates at zero external bias, offering fast response speed and zero threshold voltage. It has been proposed that the Weyl semimetal NbP and ferroelectric semiconductor GeTe are promising candidates for terahertz/infrared photodetection due to their significant current responsivity, even in the absence of external bias. We note that ferroelectrics can achieve nonvolatile manipulation of heat current via the nonlinear phonon Hall effects under electric-field control \cite{luo2023nonlinear}. 

Notable applications might arise in spintronics, where the efficiency of spin and charge conversion is predominantly governed by the spin Hall conductivity induced by Berry curvature. Different quantum numbers, such as spin and orbital, give rise to Berry curvature and BCD phenomena, known as spin-sourced BCD and orbital-sourced BCD. These distinct sources collectively demonstrate a convergence of spintronic and optoelectronic responses within a unified framework as demostrated by Lesne \textit{et al.}~\cite{lesne2023designing}. In particular, the presence of orbital-sourced Berry curvature contributes to photogalvanic currents capable of inducing spin Hall voltage in the presence of spin-sourced Berry curvature. Empirical confirmation of the nonlinear Hall effect sets the stage for realizing higher-order spin-to-charge conversions beyond linear-response theories~\cite{kozuka2021observation}. Moreover, local spin-charge interconversion can be achieved through the fabrication of tailored van der Waals heterostructures exhibiting discernible spin-orbit coupling effects~\cite{powalla2022berry}.

Spin polarization exerts a significant influence on Kerr signals, observable through angle-dependent Kerr rotation microscopy, thereby serving as tangible evidence of non-zero BCD within the sample. Additionally, phenomena such as spin-momentum locking, band tilting, and symmetry-related effects, such as Fermi surface warping give rise to spin-polarized nonlinear Nernst effects~\cite{zeng2022band}. Furthermore, the anomalous nonlinear Hall effect proves instrumental in discerning the orientation of the Neel vector and its electrically-induced manipulation via spin-orbit torques in the realm of antiferromagnetic spintronics. The orientation of the Neel vector in antiferromagnets critically hinges upon the BCD of these systems, invariably resulting in a conspicuous nonlinear anomalous Hall effect~\cite{shao2020nonlinear,wang2021intrinsic}. Recent experimental efforts have investigated the nonlinear Hall response following the integration of black phosphorus with even-layered parity-time symmetric antiferromagnet MnBi$_2$Te$_4$~\cite{gao2023quantum}. Remarkably, the non-dissipative nonlinear Hall response undergoes inversion in direction upon flipping the spins of MnBi$_2$Te$_4$. It is worthy to note that the high-harmonic spectrum exhibits intriguing characteristics in laser-driven electron dynamics within a Weyl semimetal with disrupted time-reversal symmetry \cite{bharti2022high}. Furthermore, a suitably engineered strain in phosphorene elicits a pronounced BCD compared to its pristine counterpart because of the reduced band gap~\cite{bandyopadhyay2023berry}. Moreover, the interfacing of these two-dimensional systems with antiferromagnetic or high spin-orbit coupling systems, presents a promising platform for spintronics applications~\cite{baltz2018antiferromagnetic,vzelezny2018spin}. Thus, the synergy between nonlinear transport phenomena and spintronics engenders a myriad of intriguing properties with potential utility across various applications.

In summary, we reviewed the key concepts underpinning the NLH effects, including the Berry curvature and its dipole. We discussed the analogy to nonlinear optics and summarized the intrinsic and extrinsic contributions to NLH phenomena. Leveraging symmetry indicators, we highlighted candidate materials predicted to host such physics. Up-to-date developments along the experimental front were summarized. Finally, we highlighted some of the most recent directions pushing the boundaries of this emerging area. In conclusion, we expect that many intriguing aspects of NLH and releated effects will be uncovered in the near future, and we hope that this review can motivate some of them.

\section*{Acknowledgements}

We thank S. Bhowal, M. Jain, H. R. Krishnamurthy, A. Mahajan, S. Roy, S. Saha, N. Spaldin, D. Varghese, and K. Das for valuable discussions and related collaborations. A.B. acknowledges the financial support from Indian Institute of Science IoE postdoctoral fellowship. N.B.J. acknowledges Prime Minister’s Research Fellowship for support. A.N. acknowledges support from the start-up grant (SG/MHRD-19-0001) of the Indian Institute of Science. 

\bibliography{references.bib}

\end{document}